\documentclass[aip,jcp,preprint,groupedaddress]{revtex4-2}%
\usepackage{graphicx}
\usepackage{tabularx}
\usepackage{dcolumn}
\usepackage{longtable}
\usepackage{tensor}
\usepackage{color}
\usepackage{placeins}
\usepackage{bm}
\usepackage{amsmath}
\usepackage{amsfonts}
\usepackage{amssymb}
\usepackage{float}%
\providecommand{\U}[1]{\protect\rule{.1in}{.1in}}
\begin{document}
\title{High-order geometric integrators for representation-free Ehrenfest dynamics}
\author{Seonghoon Choi}
\email{seonghoon.choi@epfl.ch}
\author{Ji\v{r}\'i Van\'i\v{c}ek}
\email{jiri.vanicek@epfl.ch}
\affiliation{Laboratory of Theoretical Physical Chemistry, Institut des Sciences et
Ing\'enierie Chimiques, Ecole Polytechnique F\'ed\'erale de Lausanne (EPFL),
CH-1015, Lausanne, Switzerland}
\date{\today}

\begin{abstract}
Ehrenfest dynamics is a useful approximation for \textit{ab initio} mixed
quantum-classical molecular dynamics that can treat electronically
nonadiabatic effects. Although a severe approximation to the exact solution of
the molecular time-dependent Schr\"{o}dinger equation, Ehrenfest dynamics is
symplectic, time-reversible, and conserves exactly the total molecular energy
as well as the norm of the electronic wavefunction. Here, we surpass apparent
complications due to the coupling of classical nuclear and quantum electronic
motions and present efficient geometric integrators for
\textquotedblleft representation-free\textquotedblright\ Ehrenfest
dynamics, which do not rely on a diabatic or adiabatic
representation of electronic states and are of arbitrary even orders
of accuracy in the time step. These numerical integrators, obtained by
symmetrically composing the second-order splitting method and exactly solving
the kinetic and potential propagation steps, are norm-conserving, symplectic,
and time-reversible regardless of the time step used. Using a nonadiabatic
simulation in the region of a conical intersection as an example, we
demonstrate that these integrators preserve the geometric properties exactly
and, if highly accurate solutions are desired, can be even more efficient than
the most popular non-geometric integrators.

\end{abstract}
\maketitle

\graphicspath{{./figures/}{C:/Users/Jiri/Dropbox/Papers/Chemistry_papers/2021/Ehrenfest_rep_indepent/figures/}}

\section{\label{sec:introduction}Introduction}

Mixed quantum-classical methods, such as the surface
hopping,\cite{Tully_Preston:1971, Tully:1990, Schmidt_Tully:2008,
Lasser_Swart:2008,Subotnik_Shenvi:2011} mean-field Ehrenfest
dynamics,\cite{Billing:1975, Billing:1976, Tully:1998, Micha:1983,
Sawada_Metiu:1985, Micha_Runge:1994, Micha:1999, Li_Frisch:2005,
Bastida_Miguel:2008, Vacher_Robb:2014} and methods based on the mixed
quantum-classical Liouville
equation\cite{Donoso_Martens:1998,Kapral_Ciccotti:1999,Shi_Geva:2004} or the
Meyer--Miller--Stock--Thoss mapping Hamiltonian,\cite{Meyer_Miller:1979,
Stock_Thoss:1997, Miller:2009, Cotton_Miller:2017,
Dunkel_Bonella:2008,Ananth_Miller:2010, Hele_Ananth:2016} remedy one of the
shortcomings of classical molecular dynamics: its inability to describe
electronically nonadiabatic processes\cite{Domcke_Yarkony:2012,
book_Takatsuka:2015, Bircher_Rothlisberger:2017} involving significantly
coupled\cite{Zimmermann_Vanicek:2010,
Zimmermann_Vanicek:2012,Zimmermann_Vanicek:2012a} states. Although a severe
approximation to the exact quantum solution,\cite{Tully:1998,
Parandekar_Tully:2006, Loaiza_Izmaylov:2018} Ehrenfest dynamics can provide a
useful first picture of nonadiabatic dynamics in some, especially strongly
coupled systems. Indeed, Ehrenfest dynamics was successfully used to describe
electron transfer,\cite{Blancafort_Robb:2005, Wang_Gou:2011, Xie_Qiang:2013,
Li_Ratner:2013, Akimov_Prezhdo:2014} nonadiabatic processes at metal
surfaces,\cite{Kirson_Ratner:1984, Kirson_Ratner:1985, Head-Gordon_Tully:1995,
Ryabinkin_Izmaylov:2017} and photochemical
processes.\cite{Topaler_Truhlar:1998, Klein_Bernardi:1998,
Gherib_Izmaylov:2015} The mean-field theory was also employed to simplify the
evaluation of the memory kernel in the generalized master equation
formalism.\cite{Markland_Kelly:2015} In addition, Ehrenfest dynamics provides
a starting point for various refined methods. For example, a multi-trajectory,
locally mean-field generalization of Ehrenfest dynamics was used to evaluate
vibronic spectra\cite{Zimmermann_Vanicek:2014} and, when combined
with the semiclassical initial value representation, can describe even
wavepacket splitting.\cite{Ananth_Miller:2007} A further
generalization, the multiconfigurational Ehrenfest
method,\cite{Shalashilin:2009,Ma_Burghardt:2018,Chen_Zhao:2021} includes
correlations between Ehrenfest trajectories. In what follows, we shall only
consider the basic, mean-field Ehrenfest method, whose validity conditions
were formulated by Bornemann \textit{et al.}\cite{Bornemann_Schutte:1996}

The coupling between nuclear and electronic dynamics complicates the numerical
integration in Ehrenfest dynamics. The widely-used two and three time step
methods\cite{Feng_Runge:1991, Micha_Runge:1994, Micha:1999, Li_Frisch:2005,
Ding_Li:2015} improve the efficiency by using different integration time steps
that account for the different time scales of nuclear and electronic motions
(see Appendix~\ref{sec:two_and_three_time_step}). However, such integration
schemes violate the geometric properties of the exact solution: the simpler,
two time step method is irreversible and neither method is symplectic (see
Fig.~\ref{fig:geom_prop_vs_two_three_step} in
Appendix~\ref{sec:two_and_three_time_step}). Almost every geometric
property\cite{book_Hairer_Wanner:2006, book_Leimkuhler_Reich:2004,
book_Lubich:2008} can, however, be preserved exactly by employing the
symplectic integrators\cite{Nettesheim_Schutte:1996} based on the splitting
method.\cite{Strang:1968, McLachlan_Quispel:2002} This splitting method is
widely applicable---so long as the Hamiltonian can be decomposed into exactly
solvable parts---and was employed to obtain symplectic integrators in many
well-known applications, including molecular quantum\cite{Feit_Steiger:1982}
and classical\cite{Verlet:1967} dynamics,
Schr\"{o}dinger--Liouville--Ehrenfest dynamics,\cite{Fang_Sparber:2018} and
the Meyer--Miller--Stock--Thoss mapping
approach.\cite{Kelly_Kapral:2012, Richardson_Thoss:2017,
Church_Ananth:2018} In particular, because the Ehrenfest method in either the
adiabatic or diabatic representation can be formulated as a special case of
the mapping method,\cite{Meyer_Miller:1979, Runeson_Richardson:2020} the
integrators developed for one of these two methods should also be applicable
to the other. Motivated by on-the-fly \textit{ab initio} applications
that employ increasingly practical real-time time-dependent electronic
structure methods,\cite{Goings_Li:2018,Li_Lopata:2020} here we present
integrators that---in contrast to the integrators\cite{Kelly_Kapral:2012,
Richardson_Thoss:2017, Church_Ananth:2018} formulated in the mapping
approach---do not rely on any particular representation of electronic states
and thus avoid the expensive construction of a truncated diabatic or adiabatic
electronic basis.

Typically, to reach the same accuracy, geometric integrators need greater
computational effort than their non-geometric
counterpart.\cite{book_Hairer_Wanner:2006, book_Leimkuhler_Reich:2004} Yet,
the efficiency of geometric integrators can be improved significantly by
employing various composition methods.\cite{book_Hairer_Wanner:2006,
book_Leimkuhler_Reich:2004, Yoshida:1990, Suzuki:1990, McLachlan:1995,
Kahan_Li:1997, Sofroniou_Spaletta:2005} Thus obtained integrators of high
orders of convergence in the time step offer the best of both worlds: they are
efficient while conserving the relevant geometric structure
exactly.\cite{book_Leimkuhler_Reich:2004}

After showing analytically, in Sec.~\ref{sec:theory}, that the high-order
geometric integrators preserve almost all of the geometric properties of
Ehrenfest dynamics, in Sec.~\ref{sec:numerical_example}, we numerically
demonstrate the efficiency and geometric properties of these integrators on a
four-dimensional extension\cite{Hader_Engel:2017, Albert_Engel:2017,
Schaupp_Engel:2019} of the Shin--Metiu model.\cite{Shin_Metiu:1995,
Shin_Metiu:1996} In this system, the first and second excited adiabatic states
are coupled significantly due to a conical intersection.
Section~\ref{sec:conclusion} concludes the paper.

\section{Theory \label{sec:theory}}

\subsection{Time-dependent Hartree approximation for the molecular
wavefunction}

Quantum evolution of a molecule is governed by the time-dependent
Schr\"{o}dinger equation (TDSE)
\begin{equation}
i\hbar\frac{d}{dt}\Psi_{t}=\mathcal{H}\Psi_{t}, \label{eq:TDSE_mol}%
\end{equation}
where $\Psi_{t}$ denotes the molecular state at time $t$ and $\mathcal{H}$ is
the molecular Hamiltonian. In general, we will denote operators acting on both
nuclei and electrons by a calligraphic font, whereas the operators acting
either only on nuclei or only on electrons will have a hat. The molecular
Hamiltonian is equal to the sum
\begin{equation}
\mathcal{H}=\hat{T}_{\mathrm{nu}}+\hat{T}_{\mathrm{el}}+\mathcal{V}
\label{eq:mol_H}%
\end{equation}
of the nuclear kinetic energy operator
\begin{equation}
\hat{T}_{\mathrm{nu}}=\frac{1}{2}\hat{P}^{T}\cdot M^{-1}\cdot\hat{P},
\label{eq:T_nuc}%
\end{equation}
electronic kinetic energy operator
\begin{equation}
\hat{T}_{\mathrm{el}}=\frac{1}{2}\hat{p}^{T}\cdot m^{-1}\cdot\hat{p},
\label{eq:T_elec}%
\end{equation}
and potential energy operator $\mathcal{V}(\hat{q},\hat{Q})$. We assume that
the nuclear position $Q$ and momentum $P$ are $D$-dimensional vectors, whereas
the electronic position $q$ and momentum $p$ are $d$-dimensional vectors. The
nuclear and electronic mass matrices, $M$ and $m$, can, in general, be real
symmetric $D\times D$ and $d\times d$ matrices, respectively.

The \emph{time-dependent Hartree} (TDH)\cite{Heller:1976, Gerber_Ratner:1982,
Gerber_Buch:1982, Bisseling_Cerjan:1987, Messina_Coalson:1989,
book_Lubich:2008} approximation is an optimal approximate solution to the
molecular TDSE~(\ref{eq:TDSE_mol}) among those in which the molecular state
can be written as the\ Hartree product%
\begin{equation}
\Psi_{t}=a_{t}\chi_{t}\psi_{t} \label{eq:TDH_mol_ansatz}%
\end{equation}
of the nuclear wavepacket $\chi_{t}$ and electronic wavepacket $\psi_{t}$; the
complex number $a_{t}$ is inserted for convenience. In the TDH approximation,
obtained by applying the Dirac-Frenkel time-dependent variational
principle\cite{Dirac:1930,book_Frenkel:1934,book_Lubich:2008} to ansatz
(\ref{eq:TDH_mol_ansatz}), the prefactor evolves as\cite{book_Lubich:2008}%
\begin{equation}
a_{t}=e^{iEt/\hbar}, \label{eq:TDH_a_t}%
\end{equation}
and the nuclear and electronic states satisfy the system%
\begin{align}
i\hbar\dot{\chi_{t}}  &  =\hat{H}_{\mathrm{nu}}\chi_{t}%
,\label{eq:TDH_mean_field_TDSE_n}\\
i\hbar\dot{\psi_{t}}  &  =\hat{H}_{\mathrm{el}}\psi_{t}
\label{eq:TDH_mean_field_TDSE_e}%
\end{align}
of coupled nonlinear Schr\"{o}dinger equations with mean-field nuclear and
electronic Hamiltonian operators
\begin{align}
\hat{H}_{\mathrm{nu}}  &  :=\langle\mathcal{H}\rangle_{\psi_{t}}=\langle
\psi_{t}|\mathcal{H}|\psi_{t}\rangle,\label{eq:TDH_mean_field_H_n}\\
\hat{H}_{\mathrm{el}}  &  :=\langle\mathcal{H}\rangle_{\chi_{t}}=\langle
\chi_{t}|\mathcal{H}|\chi_{t}\rangle. \label{eq:TDH_mean_field_H_e}%
\end{align}
The mean-field operators satisfy the obvious identity $\langle\hat
{H}_{\mathrm{nu}}\rangle_{\chi_{t}}=\langle\hat{H}_{\mathrm{el}}\rangle
_{\psi_{t}}=E$. Note that the solution expressed by Eqs.~(\ref{eq:TDH_a_t}%
)-(\ref{eq:TDH_mean_field_TDSE_e}) is unique except for an obvious gauge
freedom in redistributing the phase among $a_{t}$, $\chi_{t}$, and $\psi_{t}$.

\subsection{Mixed quantum-classical limit: Ehrenfest dynamics}

In the classical limit for nuclei, the nuclear position and momentum operators
$\hat{Q}$ and $\hat{P}$ are replaced with classical variables $Q$ and $P$.
Then, the mean-field nuclear Hamiltonian (\ref{eq:TDH_mean_field_H_n}) is no
longer an operator, but a phase space function
\begin{equation}
H_{\mathrm{nu}}(Q,P)=\langle\mathcal{H} (Q,P) \rangle_{\psi_{t}},
\label{eq:Ehrenfest_H_nuc}%
\end{equation}
where $\mathcal{H}(Q,P) = \hat{T}_{\mathrm{el}} + T_{\mathrm{nu}}(P) + \hat
{V}(Q)$, and the mean-field electronic Hamiltonian
(\ref{eq:TDH_mean_field_H_e}) becomes the molecular Hamiltonian evaluated at
the current nuclear positions and momenta:
\begin{equation}
\hat{H}_{\mathrm{el}}(Q_{t},P_{t}) = \mathcal{H}(Q_{t}, P_{t}).
\label{eq:Ehrenfest_H_e}%
\end{equation}
We thus obtain the mixed quantum-classical Ehrenfest dynamics, in which the
nuclear positions and momenta evolve according to classical Hamilton's
equations of motion with Hamiltonian $H_{\mathrm{nu}}(Q,P)$, and the
electronic state evolves according to the TDSE with a time-dependent
Hamiltonian $\hat{H}_{\mathrm{el}} (Q_{t},P_{t})$:
\begin{align}
\dot{Q}_{t}  &  =\frac{\partial H_{\mathrm{nu}}}{\partial P}(Q_{t},
P_{t}),\label{eq:Ehrenfest_q_dot}\\
\dot{P}_{t}  &  =-\frac{\partial H_{\mathrm{nu}}}{\partial Q}(Q_{t}%
,P_{t}),\label{eq:Ehrenfest_p_dot}\\
i\hbar\dot{\psi}_{t}  &  =\hat{H}_{\mathrm{el}}(Q_{t},P_{t})\psi_{t}.
\label{eq:Ehrenfest_psi_dot}%
\end{align}
Note that these three differential equations are coupled and, moreover, that
the electronic TDSE is nonlinear due to this coupling.

Equations~(\ref{eq:Ehrenfest_q_dot})--(\ref{eq:Ehrenfest_psi_dot}) can be
re-expressed as more compact Hamilton's equations
\begin{align}
\dot{q}_{\mathrm{eff},t}  &  =\frac{\partial H_{\mathrm{eff}}}{\partial
p_{\mathrm{eff}}}(q_{\mathrm{eff},t},p_{\mathrm{eff},t}%
),\label{eq:Hamilton_eq_q}\\
\dot{p}_{\mathrm{eff},t}  &  =-\frac{\partial H_{\mathrm{eff}}}{\partial
q_{\mathrm{eff}}}(q_{\mathrm{eff},t},p_{\mathrm{eff},t})
\label{eq:Hamilton_eq_p}%
\end{align}
associated with an effective mixed quantum-classical Hamiltonian
\begin{align}
H_{\mathrm{eff}}(x_{\mathrm{eff}})  &  :=\langle\hat{H}_{\mathrm{el}%
}(Q,P)\rangle_{\psi}\nonumber\\
&  =\frac{1}{2\hbar}[\langle\hat{H}_{\mathrm{el}}(Q,P)\rangle_{q_{\mathrm{\psi
}}}+\langle\hat{H}_{\mathrm{el}}(Q,P)\rangle_{p_{\mathrm{\psi}}}],
\label{eq:H_MQC}%
\end{align}
acting on an extended, effective mixed quantum-classical phase space with
coordinates $x_{\mathrm{eff}}=(q_{\mathrm{eff}},p_{\mathrm{eff}})=(Q,q_{\psi
},P,p_{\psi})$. The \textquotedblleft quantum\textquotedblright\ Darboux
coordinates $(q_{\psi},p_{\psi})$ consist of the real and imaginary part of
the electronic wavefunction in position representation: $q_{\psi}%
:=\sqrt{2\hbar}\mathrm{Re}\psi(q)$ and $p_{\psi}:=\sqrt{2\hbar}\mathrm{Im}%
\psi(q)$ (we omit the dependence of $q_{\psi}$ and $p_{\psi}$ on $q$ for brevity).

In general, $q_{\psi}$ and $p_{\psi}$ are real functions in an
infinite-dimensional space; therefore, the \textquotedblleft
quantum\textquotedblright\ part
\begin{align}
\dot{q}_{\psi,t}  &  =\frac{\delta H_{\mathrm{eff}}}{\delta p_{\psi}%
}(q_{\mathrm{eff,t}},p_{\mathrm{eff},t}),\label{eq:Ehrenfest_qm_part_q_psi}\\
\dot{p}_{\psi,t}  &  =-\frac{\delta H_{\mathrm{eff}}}{\delta q_{\psi}%
}(q_{\mathrm{eff,t}},p_{\mathrm{eff},t}) \label{eq:Ehrenfest_qm_part_p_psi}%
\end{align}
of Eqs.~(\ref{eq:Hamilton_eq_q}) and (\ref{eq:Hamilton_eq_p}), in fact,
involves partial functional derivatives:\cite{book_Marsden_Ratiu:1999}
\begin{align}
\frac{\delta H_{\mathrm{eff}}}{\delta q_{\psi}}  &  =\frac{1}{2\hbar}%
\frac{\delta}{\delta q_{\psi}}\int[q_{\psi}\hat{H}_{\mathrm{el}}(Q,P)q_{\psi
}+p_{\psi}\hat{H}_{\mathrm{el}}(Q,P)p_{\psi}]dq\nonumber\\
&  =\hbar^{-1}\hat{H}_{\mathrm{el}}(Q,P)q_{\psi},\label{eq:func_deriv_H_eff_q}%
\\
\frac{\delta H_{\mathrm{eff}}}{\delta p_{\psi}}  &  =\frac{1}{2\hbar}%
\frac{\delta}{\delta p_{\psi}}\int[q_{\psi}\hat{H}_{\mathrm{el}}(Q,P)q_{\psi
}+p_{\psi}\hat{H}_{\mathrm{el}}(Q,P)p_{\psi}]dq\nonumber\\
&  =\hbar^{-1}\hat{H}_{\mathrm{el}}(Q,P)p_{\psi}.
\label{eq:func_deriv_H_eff_p}%
\end{align}
Substituting Eqs.~(\ref{eq:func_deriv_H_eff_q}) and
(\ref{eq:func_deriv_H_eff_p}) into Hamilton's
equations~(\ref{eq:Ehrenfest_qm_part_q_psi}) and
(\ref{eq:Ehrenfest_qm_part_p_psi}) recovers the
TDSE~(\ref{eq:Ehrenfest_psi_dot}) for the electronic wavefunction.

In practical calculations, $q_{\psi}$ and $p_{\psi}$ are usually represented
in a finite basis or on a grid as $N$-dimensional vectors, where $N$ is either
the size of the basis or number of grid points. In such cases, functional
derivatives~(\ref{eq:func_deriv_H_eff_q}) and (\ref{eq:func_deriv_H_eff_p})
reduce to partial derivatives
\begin{align}
\frac{\partial H_{\mathrm{eff}}}{\partial q_{\psi}}  &  =\frac{1}{2\hbar}%
\frac{\partial}{\partial q_{\psi}}[q_{\psi}^{T}H_{\mathrm{el}}(Q,P)q_{\psi
}+p_{\psi}^{T}H_{\mathrm{el}}(Q,P)p_{\psi}]\nonumber\\
&  =\hbar^{-1}H_{\mathrm{el}}(Q,P)q_{\psi},\label{eq:partial_deriv_H_eff_q}\\
\frac{\partial H_{\mathrm{eff}}}{\partial p_{\psi}}  &  =\frac{1}{2\hbar}%
\frac{\partial}{\partial p_{\psi}}[q_{\psi}^{T}H_{\mathrm{el}}(Q,P)q_{\psi
}+p_{\psi}^{T}H_{\mathrm{el}}(Q,P)p_{\psi}]\nonumber\\
&  =\hbar^{-1}H_{\mathrm{el}}(Q,P)p_{\psi}, \label{eq:partial_deriv_H_eff_p}%
\end{align}
where $H_{\mathrm{el}}(Q,P)$ is an $N\times N$ matrix representation of
operator $\hat{H}_{\mathrm{el}}(Q,P)$.

\subsection{Geometric properties \label{subsec:geom_prop}}

\subsubsection{Norm conservation}

Ehrenfest dynamics conserves the norm
\begin{equation}
\Vert\psi_{t}\Vert:=\langle\psi_{t}|\psi_{t}\rangle^{1/2} \label{eq:wf_norm}%
\end{equation}
of the electronic wavefunction because
\begin{align}
\frac{d}{dt}\Vert\psi_{t}\Vert^{2}  &  =\langle\dot{\psi}_{t}|\psi_{t}%
\rangle+\langle\psi_{t}|\dot{\psi}_{t}\rangle\nonumber\\
&  =\frac{i}{\hbar}[\langle\hat{H}_{\mathrm{el}}(Q_{t},P_{t})\rangle_{\psi
_{t}}-\langle\hat{H}_{\mathrm{el}}(Q_{t},P_{t})\rangle_{\psi_{t}}]=0,
\label{eq:Ehrenfest_norm_conservation}%
\end{align}
where we used Eq.~(\ref{eq:Ehrenfest_psi_dot}) and the hermiticity of $\hat
{H}_{\mathrm{el}}(Q_{t},P_{t})$.

\subsubsection{Energy conservation}

The total energy $E=H_{\mathrm{nu}}(Q_{t},P_{t})=\langle\hat{H}_{\mathrm{el}%
}(Q_{t},P_{t})\rangle_{\psi_{t}}$ of the system is conserved, in general, by
the time-dependent variational principle and, in particular, by the TDH
approximation. However, because we have also taken the mixed quantum-classical
limit, let us verify the conservation of energy explicitly:
\begin{align}
\frac{dE}{dt}  &  =\langle\dot{\psi_{t}}|\hat{H}_{\mathrm{el}}(Q_{t}%
,P_{t})|\psi_{t}\rangle+\langle\psi_{t}|\hat{H}_{\mathrm{el}}(Q_{t}%
,P_{t})|\dot{\psi}_{t}\rangle\nonumber\\
&  \qquad+\dot{Q}_{t}^{T}\cdot\frac{\partial H_{\mathrm{nu}}}{\partial
Q}(Q_{t},P_{t})+\dot{P}_{t}^{T}\cdot\frac{\partial H_{\mathrm{nu}}}{\partial
P}(Q_{t},P_{t})\nonumber\\
&  =i\hbar\lbrack\langle\dot{\psi}_{t}|\dot{\psi}_{t}\rangle-\langle\dot{\psi
}_{t}|\dot{\psi}_{t}\rangle]-\dot{Q}_{t}^{T}\cdot\dot{P}_{t}+\dot{P}_{t}%
^{T}\cdot\dot{Q}_{t}=0,
\end{align}
where we used the hermiticity of $\hat{H}_{\mathrm{el}}(Q_{t},P_{t})$ and
Eqs.~(\ref{eq:Ehrenfest_q_dot})--(\ref{eq:Ehrenfest_psi_dot}). The energy
conservation also follows directly from the effective Hamiltonian structure:
\begin{align}
\frac{dE}{dt}  &  =\frac{d}{dt}H_{\mathrm{eff}}(q_{\mathrm{eff},t}%
,p_{\mathrm{eff},t})\nonumber\\
&  =\dot{q}_{\mathrm{eff},t}^{T}\frac{\partial H_{\mathrm{eff}}}{\partial
q_{\mathrm{eff}}}(q_{\mathrm{eff},t},p_{\mathrm{eff},t})+\dot{p}%
_{\mathrm{eff},t}^{T}\frac{\partial H_{\mathrm{eff}}}{\partial p_{\mathrm{eff}%
}}(q_{\mathrm{eff},t},p_{\mathrm{eff},t})\nonumber\\
&  =-\dot{q}_{\mathrm{eff},t}^{T}\dot{p}_{\mathrm{eff},t}+\dot{p}%
_{\mathrm{eff},t}^{T}\dot{q}_{\mathrm{eff},t}=0,
\label{eq:Ehrenfest_energy_conservation}%
\end{align}
where we used Eqs.~(\ref{eq:Hamilton_eq_q}) and (\ref{eq:Hamilton_eq_p}).

\subsubsection{Symplecticity}

The effective, mixed quantum-classical symplectic two-form
\begin{equation}
\omega_{\mathrm{eff}}:=dq_{\mathrm{eff}}\wedge dp_{\mathrm{eff}}%
=\omega_{\mathrm{cl}}+\omega_{\mathrm{qm}} \label{eq:mqc_two_form}%
\end{equation}
is a sum of the classical (cl) canonical two-form $\omega_{\mathrm{cl}%
}:=dQ\wedge dP$ and the quantum (qm) canonical two-form $\omega_{\mathrm{qm}%
}:=dq_{\psi}\wedge dp_{\psi}$, which acts on states $\psi_{1}$ and $\psi_{2}$
as $\omega_{\text{qm}}(\psi_{1},\psi_{2})=2\hbar\mathrm{Im}\langle\psi
_{1}|\psi_{2}\rangle$ (see Appendix~\ref{sec:qm_symplec} and
Refs.~\onlinecite{book_Lubich:2008, book_Marsden_Ratiu:1999, Ohsawa_Leok:2013}).
Let $\Phi_{H_{\text{eff}},t}:x_{\text{eff,}0}\mapsto x_{\text{eff,}t}$ denote
the Hamiltonian flow of $H_{\text{eff}}$. The stability (or Jacobian) matrix
$M_{t}$ of the Hamiltonian flow $\Phi_{H,t}$ is a symplectic matrix. While
this holds in general,\cite{book_Abraham_Marsden:1978,
book_Leimkuhler_Reich:2004} we show it explicitly for our case in
Appendix~\ref{sec:Ham_flow_symplec}. As a result, Ehrenfest dynamics conserves
the symplectic two-form $\omega_{\mathrm{eff}}$ from
Eq.~(\ref{eq:mqc_two_form}) (see Appendix~\ref{sec:Ham_flow_symplec}).

\subsubsection{Time reversibility}

An involution is a mapping $S$ that is its own inverse, i.e., $S(S(x))=x$. We
will consider the involution
\begin{equation}
S=%
\begin{pmatrix}
I_{D+N} & 0\\
0 & -I_{D+N}%
\end{pmatrix}
\label{eq:t_rev_S_mat}%
\end{equation}
that changes the sign of the nuclear momenta and conjugates the electronic
wavefunction in position representation (i.e., changes the sign of $p_{\psi}%
$). Following Ref.~\onlinecite{book_Leimkuhler_Reich:2004}, we call a flow
$\Phi_{t}$ time-reversible under a general involution $S$ if it satisfies
\begin{equation}
S\Phi_{t}[S\Phi_{t}(x_{\mathrm{eff}})]=x_{\mathrm{eff}}.
\label{eq:t_rev_condition}%
\end{equation}
Because effective Hamiltonian $H_{\mathrm{eff}}$ is an even function in
$p_{\mathrm{eff}}$, i.e., $H_{\mathrm{eff}}(x_{\mathrm{eff}})=H_{\mathrm{eff}%
}(Sx_{\mathrm{eff}})$, its Hamiltonian flow
satisfies\cite{book_Leimkuhler_Reich:2004}
\begin{equation}
\Phi_{H_{\mathrm{eff}},t}(x_{\mathrm{eff}})=S\Phi_{H_{\mathrm{eff}}%
,-t}(Sx_{\mathrm{eff}}). \label{eq:t_rev_condition_equiv}%
\end{equation}
Since $S^{-1}=S$ and, by definition, any flow is symmetric (i.e., $\Phi
_{-t}=\Phi_{t}^{-1}$),\cite{book_Hairer_Wanner:2006,
book_Leimkuhler_Reich:2004} the satisfaction of
Eq.~(\ref{eq:t_rev_condition_equiv}) implies the satisfaction of the time
reversibility condition~(\ref{eq:t_rev_condition}).

\subsection{Geometric Integrators \label{subsec:geometric_integrators}}

As in the split-operator algorithm\cite{Feit_Steiger:1982,
Wehrle_Vanicek:2011,Roulet_Vanicek:2019, Choi_Vanicek:2019a} for the TDSE or
in the Verlet algorithm\cite{Verlet:1967} for Hamilton's equations of motion,
we can obtain a symmetric potential-kinetic-potential (VTV) algorithm of the
second order in the time step $\Delta t$ by using the Strang
splitting\cite{Strang:1968} and performing, in sequence, potential propagation
for time $\Delta t/2$, kinetic propagation for $\Delta t$, and potential
propagation for $\Delta t/2$. The second-order kinetic-potential-kinetic (TVT)
algorithm is obtained similarly, by exchanging the potential and kinetic
propagations. Either of the two second-order algorithms can be symmetrically
composed\cite{book_Hairer_Wanner:2006,Choi_Vanicek:2019} to obtain an
algorithm of an arbitrary even order of accuracy in $\Delta t$. This is
achieved by using the recursive triple-jump\cite{Yoshida:1990,Suzuki:1990} or
Suzuki-fractal\cite{Suzuki:1990} composition schemes, or with a more efficient
scheme specific to each order (which we shall call \textquotedblleft
optimal\textquotedblright).\cite{Suzuki:1990, Kahan_Li:1997,
Sofroniou_Spaletta:2005} We, therefore, only need to present the analytical
solutions of the kinetic and potential propagation steps for arbitrary times
$t$.

During the kinetic propagation, the Hamiltonian reduces to
\begin{equation}
\hat{H}(Q,P)=\hat{T}_{\mathrm{el}}+T_{\mathrm{nu}}(P), \label{eq:T_prop_H}%
\end{equation}
and the equations of motion~(\ref{eq:Ehrenfest_q_dot}%
)--(\ref{eq:Ehrenfest_psi_dot}) become
\begin{align}
\dot{Q_{t}}  &  =M^{-1}\cdot P_{t},\label{eq:ehrenfest_kinetic_q}\\
\dot{P_{t}}  &  =0,\label{eq:ehrenfest_kinetic_p}\\
i\hbar\dot{\psi_{t}}  &  =[\hat{T}_{\mathrm{el}}+T_{\mathrm{nu}}(P_{t}%
)]\psi_{t}, \label{eq:ehrenfest_kinetic_wf}%
\end{align}
which are equivalent to Hamilton's equations~(\ref{eq:Hamilton_eq_q}) and
(\ref{eq:Hamilton_eq_p}) with $H_{\mathrm{eff}}=\langle\hat{T}_{\mathrm{el}%
}+T_{\mathrm{nu}}(P)\rangle_{\psi}$. Because nuclear momenta $P_{t}$ do not
evolve during the kinetic propagation, Eqs.~(\ref{eq:ehrenfest_kinetic_q}%
)--(\ref{eq:ehrenfest_kinetic_wf}) can be solved analytically to obtain
\begin{align}
Q_{t}  &  =Q_{0}+tM^{-1}\cdot P_{0},\label{eq:ehrenfest_kinetic_sol_q}\\
P_{t}  &  =P_{0},\label{eq:ehrenfest_kinetic_sol_p}\\
\psi_{t}  &  =e^{-it[\hat{T}_{\mathrm{el}}+T_{\mathrm{nu}}(P_{0})]/\hbar}%
\psi_{0}. \label{eq:ehrenfest_kinetic_sol_wf}%
\end{align}
As $\hat{T}_{\mathrm{el}}=T(\hat{p})$, Eq.~(\ref{eq:ehrenfest_kinetic_sol_wf})
is easily evaluated in momentum representation.

During the potential propagation, the Hamiltonian reduces to
\begin{equation}
\hat{H}(Q,P)=\hat{V}(Q), \label{eq:V_prop_H}%
\end{equation}
and the equations of motion~(\ref{eq:Ehrenfest_q_dot}%
)--(\ref{eq:Ehrenfest_psi_dot}) become
\begin{align}
\dot{Q_{t}}  &  =0,\label{eq:ehrenfest_potential_q}\\
\dot{P_{t}}  &  =-\langle\hat{V}^{\prime}(Q_{t})\rangle_{\psi_{t}%
},\label{eq:ehrenfest_potential_p}\\
i\hbar\dot{\psi_{t}}  &  =\hat{V}(Q_{t})\psi_{t},
\label{eq:ehrenfest_potential_wf}%
\end{align}
which are equivalent to Hamilton's equations~(\ref{eq:Hamilton_eq_q}) and
(\ref{eq:Hamilton_eq_p}) with $H_{\mathrm{eff}}=\langle\hat{V}(Q)\rangle
_{\psi}$. Because nuclear positions $Q_{t}$ do not evolve during the potential
propagation, one can replace $Q_{t}$ with $Q_{0}$ in
Eqs.~(\ref{eq:ehrenfest_potential_p}) and (\ref{eq:ehrenfest_potential_wf}).
Even after the substitution of $Q_{t}=Q_{0}$,
Eq.~(\ref{eq:ehrenfest_potential_p}) seems hard to solve due to an apparent
coupling to Eq.~(\ref{eq:ehrenfest_potential_wf}). However, this coupling can
be removed by noting that
\begin{align}
\langle\hat{V}^{\prime}(Q_{0})\rangle_{\psi_{t}}  &  =\langle\psi
_{0}|e^{it\hat{V}(Q_{0})/\hbar}\hat{V}^{\prime}(Q_{0})e^{-it\hat{V}%
(Q_{0})/\hbar}|\psi_{0}\rangle\nonumber\\
&  =\langle\psi_{0}|e^{it\hat{V}(Q_{0})/\hbar}e^{-it\hat{V}(Q_{0})/\hbar}%
\hat{V}^{\prime}(Q_{0})|\psi_{0}\rangle\nonumber\\
&  =\langle\hat{V}^{\prime}(Q_{0})\rangle_{\psi_{0}}.
\label{eq:V_wf_t_equals_V_wf_0}%
\end{align}
As a result, Eqs.~(\ref{eq:ehrenfest_potential_q}%
)--(\ref{eq:ehrenfest_potential_wf}) can be solved analytically to obtain
\begin{align}
Q_{t}  &  =Q_{0},\label{eq:ehrenfest_potential_sol_q}\\
P_{t}  &  =P_{0}-t\langle\hat{V}^{\prime}(Q_{0})\rangle_{\psi_{0}%
},\label{eq:ehrenfest_potential_sol_p}\\
\psi_{t}  &  =e^{-it\hat{V}(Q_{0})/\hbar}\psi_{0}.
\label{eq:ehrenfest_potential_sol_wf}%
\end{align}
As $\hat{V}(Q)=V(\hat{q},Q)$, Eq.~(\ref{eq:ehrenfest_potential_sol_wf}) is
easily evaluated in position representation.

\subsection{Geometric properties of the geometric integrator
\label{subsec:geom_prop_geom_int}}

Like all other integrators obtained with the splitting method, in the proposed
algorithm, each potential or kinetic step of the Ehrenfest dynamics is solved
exactly, and each of these steps has all the geometric properties of the exact
solution. The second-order Strang splitting method, composed from two exact
flows, or any of its symmetric compositions preserves all the listed geometric
properties except the conservation of energy (see, e.g.,
Refs.~\onlinecite{book_Leimkuhler_Reich:2004,book_Hairer_Wanner:2006} or
Refs.~\onlinecite{Roulet_Vanicek:2019, Choi_Vanicek:2019a, Choi_Vanicek:2019}
and references therein).

\section{Numerical example \label{sec:numerical_example}}

We use a low-dimensional model that is solvable \textquotedblleft
numerically\textquotedblright\ exactly to demonstrate the geometric and
convergence properties of the presented integrators. Since the high efficiency
and geometric properties of these integrators are most meaningful when the
mean-field Ehrenfest approximation is valid, we have chosen the system and
initial state carefully so that numerically converged Ehrenfest and exact
quantum simulations yield similar results. At the same time, we have ensured
that the resulting nonadiabatic simulation describes a realistic light-induced
excitation. (See references in Sec.~\ref{sec:introduction} for
higher-dimensional examples where the Ehrenfest approximation was employed
successfully.)

The four-dimensional
extension\cite{Hader_Engel:2017,Albert_Engel:2017,Schaupp_Engel:2019} of the
Shin--Metiu model\cite{Shin_Metiu:1995,Shin_Metiu:1996} consists of an
interacting electron and proton, both moving in two spatial dimensions and
feeling an additional field of two fixed protons (all three protons are
distinguishable). The four-dimensional ($D=2$, $d=2$) model Hamiltonian is of
form~(\ref{eq:mol_H}) with
\begin{align}
\mathcal{V}  &  =V_{\mathrm{quartic}}(\hat{Q})+V^{en}(|\hat{q}-Q_{a}%
|)+V^{en}(|\hat{q}-Q_{b}|)\nonumber\\
&  \qquad+V^{en}(|\hat{q}-\hat{Q}|)+V^{nn}(|\hat{Q}-Q_{a}|)\nonumber\\
&  \qquad+V^{nn}(|\hat{Q}-Q_{b}|)+V^{nn}(|Q_{a}-Q_{b}|),
\label{eq:two_d_D_shin_metiu_pot}%
\end{align}
where $Q_{a}=(-L/2,0)$ and $Q_{b}=(L/2,0)$ are the positions of the two fixed
protons, and
\begin{align}
V^{en}(\xi)  &  =-1/\sqrt{a+\xi^{2}}\label{eq:coulomb_en}\\
V^{nn}(\xi)  &  =1/\sqrt{b+\xi^{2}} \label{eq:coulomb_nn}%
\end{align}
are attractive and repulsive regularized Coulomb potentials; following
Ref.~\onlinecite{Hader_Engel:2017}, we take $L=4\sqrt{3}/5$ a.u.,
$a=0.5\ (\mathrm{a.u.})^{2}$, and $b=10\ (\mathrm{a.u.})^{2}$. Quartic
potential $V_{\mathrm{quartic}}(Q)=(|Q|/Q_{c})^{4}$ with $Q_{c}=3.5$ a.u.
ensures that the system remains bound.

For the dynamics, we considered the initial state
\begin{equation}
\Psi_{0}(q,Q)=\chi_{\mathrm{gwp}}(Q-Q_{0})\varphi_{2}(q;Q),
\label{eq:init_state}%
\end{equation}
where $\varphi_{i}$ is the $i$th excited adiabatic electronic state, and
\begin{equation}
\chi_{\mathrm{gwp}}(Q)=\frac{1}{\sqrt{\pi\hbar\sigma^{2}}}e^{-Q^{2}%
/(2\hbar\sigma^{2})} \label{eq:shin_metiu_init_state}%
\end{equation}
is the ground vibrational eigenstate of a harmonic fit to the ground
electronic state; here, $\sigma=0.24$ a.u. The displacement of the initial
wavepacket by $Q_{0}=(0.5\ \mathrm{a.u.},1.5\ \mathrm{a.u.})$ from the ground
state equilibrium is motivated by the displaced excitation of molecules:
Suppose state $\varphi_{2}$ is dark; then a wavepacket may reach it at a
nuclear geometry that is not the ground state equilibrium via an intersection
with a bright state. To obtain $\varphi_{2}(q;Q)$, we solved the electronic
time-independent Schr\"{o}dinger equation
\begin{equation}
\lbrack\hat{T}_{\mathrm{el}}+\hat{V}(Q)]\varphi_{i}(Q)=E_{i}(Q)\varphi_{i}(Q),
\label{eq:electronic_tise}%
\end{equation}
where $\hat{T}_{\mathrm{el}}=T(\hat{p})$ and $\hat{V}%
(Q)=V(\hat{q},Q)$ are operators acting on electrons; in position
representation, Eq.~(\ref{eq:electronic_tise}) takes a more familiar form
\begin{equation}
\left[  -\frac{\hbar^{2}}{2m}\frac{\partial^{2}}{\partial q^{2}}%
+V(q;Q)\right]  \varphi_{i}(q;Q)=E_{i}(Q)\varphi_{i}(q;Q).
\label{eq:electronic_tise_pos_rep}%
\end{equation}
Section~S1 of the supplementary material describes the method we employed to
solve this equation.

\begin{figure}
\includegraphics[scale=1.0]{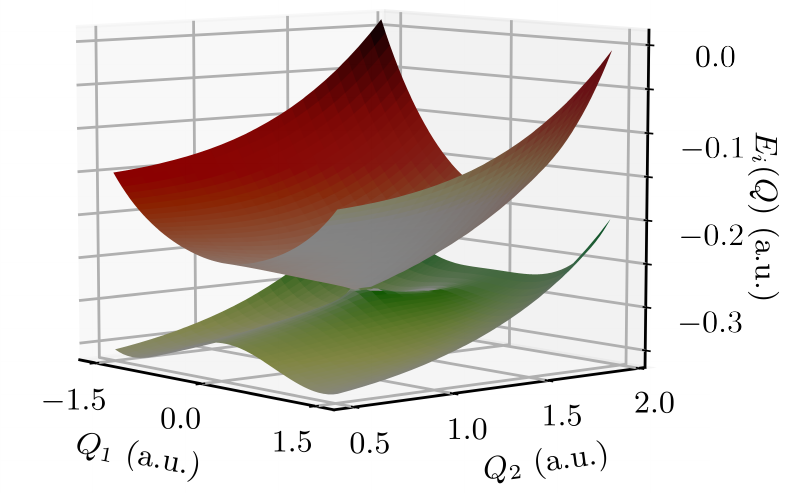}
\caption{Potential energy surfaces in the vicinity of the conical intersection at
$Q=Q_{\mathrm{CI}}=(0, 1.2 \ \mathrm{a.u.})$ in the model system from Sec.~\ref{sec:numerical_example}.
Energies $E_{i}(Q) := \langle \varphi_{i}(Q) | \hat{T}_{\mathrm{el}} + \hat{V}(Q) | \varphi_{i}(Q) \rangle$ of the first
($i=1$) and second ($i=2$)  excited adiabatic electronic states $\varphi_{i}(Q)$ are shown in green and red, respectively.}\label{fig:pes3d}%

\end{figure}

Because our approach does not rely on a specific electronic basis (such as the
basis of adiabatic or diabatic electronic states) to represent the molecular
wavepacket, the initial state can be a general function of $q$ and $Q$. To be
specific, however, we chose to start the dynamics from a single excited
adiabatic electronic state (here $\varphi_{2}$), which is the most common
choice in the literature studying nonadiabatic dynamics following a
light-induced excitation.\cite{book_Nakamura:2012, book_Mukamel:1999,
book_Domcke_Koppel:2004, book_Takatsuka:2015, book_Heller:2018} In the model
described by Eqs.~(\ref{eq:two_d_D_shin_metiu_pot})--(\ref{eq:coulomb_nn}),
the second excited adiabatic state $\varphi_{2}$ is, indeed, significantly
coupled to the first excited state $\varphi_{1}$ by a conical intersection
depicted in Fig.~\ref{fig:pes3d}.

\begin{figure}
\includegraphics[scale=1.0]{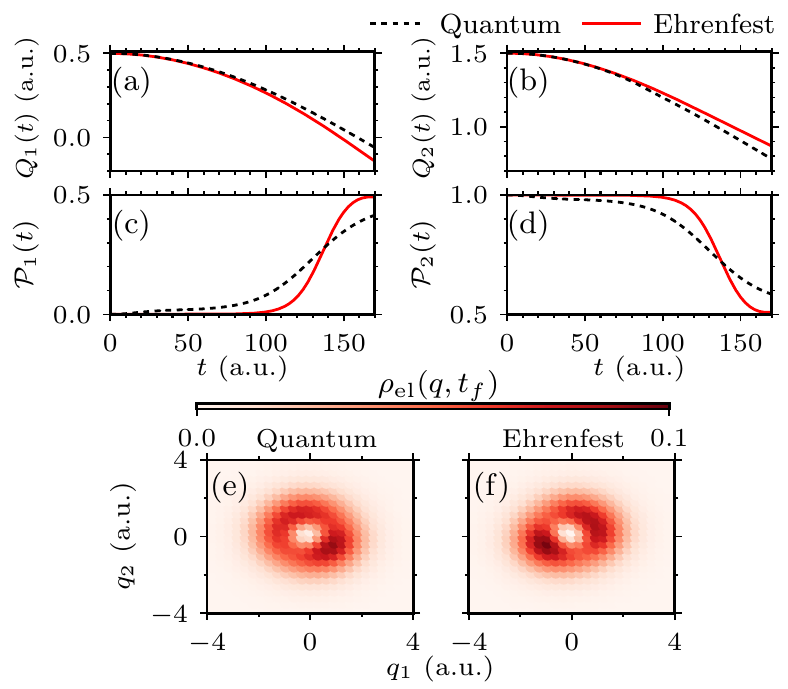}
\caption{Comparison of Ehrenfest dynamics with the exact quantum dynamics. (a)--(d): Time dependence of
nuclear position $Q_{n}(t)$ [$n=1$ in panel~(a), $n=2$ in panel~(b)] and of the first and second excited adiabatic
populations $\mathcal{P}_{i}(t)$ [$i=1$ in panel~(c), $i=2$ in panel~(d)]. (e) and (f): Electronic densities at the final
time $t=t_{f}$ obtained from quantum dynamics [panel~(e)] and Ehrenfest dynamics [panel~(f)]. }\label{fig:qm_vs_Ehrenfest}%

\end{figure}

In Fig.~\ref{fig:qm_vs_Ehrenfest}, we compare the exact quantum dynamics
$\Psi_{t}=\exp{(-it\mathcal{H}/\hbar)}\Psi_{0}$ with Ehrenfest dynamics
$x_{\mathrm{eff},t}=\Phi_{H_{\mathrm{eff}},t}(x_{\mathrm{eff},0})$. The
initial state of the system is $\left(  Q_{t},P_{t},\psi_{t}\right)
|_{t=0}=(Q_{0},0,\varphi_{2}(Q_{0}))$ and the corresponding initial mixed
quantum-classical phase space point
\begin{equation}
x_{\mathrm{eff},0}=(Q_{0},\sqrt{2\hbar}\mathrm{Re}\varphi_{2}(q;Q_{0}),0,0)
\label{eq:init_mixed_qm_cl_pnt}%
\end{equation}
 can be thought of as state~(\ref{eq:init_state}) with an infinitesimally
narrow Gaussian wavepacket; the fourth component in
Eq.~(\ref{eq:init_mixed_qm_cl_pnt}) is zero because the state $\varphi
_{2}(q;Q_{0})$, in position representation, is purely real: $\mathrm{Im}%
\varphi_{2}(q;Q_{0})=0$. We compare three observables: nuclear position
$Q(t)$, adiabatic population $\mathcal{P}_{i}(t)$, and electronic density
$\rho_{\mathrm{el}}(q,t)$. In quantum dynamics, they are obtained from the
full wavefunction $\Psi_{t}$ as\cite{Hader_Engel:2017,Schaupp_Engel:2019}
\begin{align}
Q(t)  &  =\langle\hat{Q}\rangle_{\Psi_{t}},\label{eq:position_qm}\\
\mathcal{P}_{i}(t)  &  =\langle\hat{\mathcal{P}}_{i}\rangle_{\Psi_{t}%
},\label{eq:adiab_pop_qm}\\
\rho_{\mathrm{el}}(q,t)  &  =\int dQ|\Psi_{t}(q,Q)|^{2}.
\label{eq:rho_elec_qm}%
\end{align}
To find population $\mathcal{P}_{i}(t)$ from Eq.~(\ref{eq:adiab_pop_qm}), we
computed the expectation value of the population operator $\hat{\mathcal{P}%
}_{i}:=|\varphi_{i}\rangle\langle\varphi_{i}|$ in position
representation:\cite{Schaupp_Engel:2019}
\begin{equation}
\langle\hat{\mathcal{P}}_{i}\rangle_{\Psi_{t}}=\int\left\vert \int\varphi
_{i}(q;Q)^{\ast}\Psi_{t}(q,Q)dq\right\vert ^{2}dQ.
\end{equation}
In Ehrenfest dynamics, the nuclear position $Q(t)$ is simply the current
position $Q_{t}$ of the trajectory, whereas the adiabatic population
$\mathcal{P}_{i}(t)=|\langle\varphi_{i}(Q_{t})|\psi_{t}\rangle|^{2}$ and
electronic density $\rho_{\mathrm{el}}(q,t)=|\psi_{t}(q)|^{2}$ depend on the
electronic wavefunction $\psi_{t}$.

Figure~\ref{fig:qm_vs_Ehrenfest} shows that during the considered time
interval $t\in\lbrack0,t_{f}]$ with $t_{f}=170$ a.u., Ehrenfest dynamics
yields qualitatively correct results. In particular, the nuclear motion
towards the conical intersection at $Q=Q_{\mathrm{CI}}=(0,1.2\ \mathrm{a.u.})$
[panels~(a) and (b)] and the resulting population transfer from the initial
second excited to the first excited state [panels~(c) and (d)] are well
described by Ehrenfest dynamics. The electronic densities obtained from the
exact quantum and Ehrenfest dynamics [panels~(e) and (f)] at the final time
$t=t_{f}$ are also very similar. The mean-field Ehrenfest
approximation works well because the nuclear density remains localized (not
shown) and the electronic density is almost stationary.

In the following, we demonstrate the geometric properties and high
efficiency of high-order geometric integrators (from
Sec.~\ref{subsec:geometric_integrators}). Owing to the low electronic
dimensionality ($d=2$) of the employed model, we could ensure that the
numerical errors due to the representation of the electronic wavefunction
$\psi_{t}(q)$ were negligible in comparison with the time propagation errors:
The wavefunction was represented, with high accuracy, on a uniform grid (see
Sec.~S1 of the supplementary material for computational details). This
approach, however, would be too computationally demanding in practical,
higher-dimensional (i.e., larger $d$) simulations. Instead, for such
simulations, one of the real-time time-dependent electronic structure
methods,\cite{Goings_Li:2018,Li_Lopata:2020} such as real-time time-dependent
Hartree--Fock (TDHF)\cite{Micha_Runge:1994, Li_Frisch:2005, Li_Schlegel:2005}
and real-time time-dependent density functional theory
(TDDFT),\cite{Theilhaber:1992, Yabana_Bertsch:1996, Castro_Rubio:2004,
Isborn_Tully:2007, Miyamoto_Tomanek:2006, Meng_Kaxiras:2008,
Andrade_Rubio:2009, Liang_Levis:2010} should be employed. In particular,
because there are existing implementations of real-time TDDFT using the
split-operator algorithms for the propagation of Kohn--Sham
orbitals,\cite{Castro_Rubio:2004a, book_Marques_Gross:2003,
Marques_Rubio:2003, Andrade_Marques:2012, Tancogne-Dejean_Rubio:2020} it
should be straightforward to employ the presented integrators for
TDDFT-Ehrenfest simulations. However, the exact efficiency of the high-order
geometric integrators when applied to realistic TDDFT-Ehrenfest simulations is
hard to predict and is outside the scope of this study.

\begin{figure}
\includegraphics[scale=1.0]{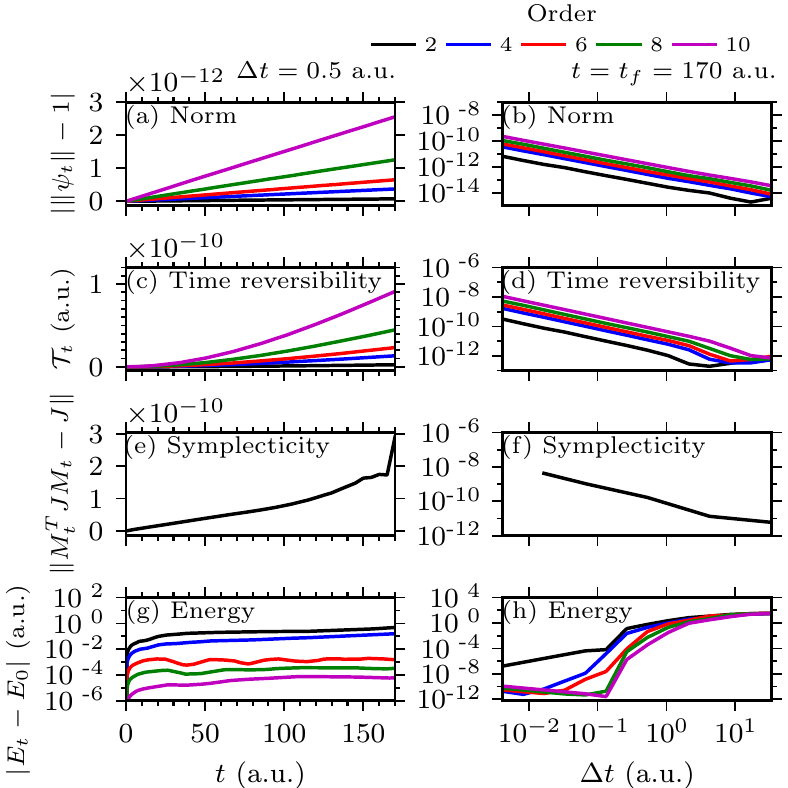}
\caption{Conservation of geometric properties by the geometric integrators presented in
Sec.~\ref{subsec:geometric_integrators}. (a) and (b): Norm of the electronic wavefunction. (c) and (d): Time
reversibility [Eq.~(\ref{eq:time_reversibility})]. (e) and (f): Symplecticity [Eq.~(\ref{eq:symplecticity})]. (g) and (h): Energy. Both the time dependence
of the geometric properties for a fixed time step $\Delta t = 0.5$ a.u. [left-hand side panels~(a), (c), (e), and (g)] and
the geometric properties at the final time, $t=t_{f}=170$ a.u., as functions of $\Delta t$ [right-hand side
panels~(b), (d), (f), and (h)] are shown. The costly numerical propagation of stability matrix $M_{t}$ is done
separately from the main Ehrenfest dynamics (see Appendix~\ref{sec:Num_prop_stab_mat}). Due to prohibitive
computational cost, only the elementary second-order method is presented in panels~(e) and (f); however, all of its
compositions are symplectic regardless of the time step (as justified in Sec.~\ref{subsec:geom_prop_geom_int}).
Initial energy of the system is $E_{0}=-0.2$ a.u. To avoid clutter, only the higher-order integrators obtained using the optimal
composition schemes are shown (the Suzuki-fractal scheme is the optimal fourth-order
scheme\cite{Choi_Vanicek:2019}).}\label{fig:geom_prop}
\end{figure}

In Fig.~\ref{fig:geom_prop}, we demonstrate the geometric properties of the
presented integrators (in all figures, we omit the results of the TVT
algorithm and its compositions because they are nearly identical to the
corresponding results for the VTV algorithm). The figure shows that the norm
of the electronic wavefunction [panels~(a) and (b)], time reversibility
[panels~(c) and (d)], and symplecticity [panels~(e) and (f)] are conserved as
functions of time [for a fixed time step $\Delta t=0.5$ a.u., panels~(a), (c),
and (e)] and regardless of the time step $\Delta t$ used [for a fixed final
time $t_{f}=170$ a.u., panels~(b), (d), and (f)]. We check the symplecticity
of stability matrix $M_{t}$ by measuring the Frobenius distance
\begin{equation}
d_{t}=\Vert M_{t}^{T}JM_{t}-J\Vert\label{eq:symplecticity}%
\end{equation}
of $M_{t}^{T}JM_{t}$ from $J$ (see Appendix~\ref{sec:Ham_flow_symplec}).
Here,
\begin{equation}
J:=%
\begin{pmatrix}
0 & -I_{D+N}\\
I_{D+N} & 0
\end{pmatrix}
\end{equation}
is the \emph{standard symplectic matrix}, and the Frobenius norm is defined as
$\Vert A\Vert:=\langle A,A\rangle^{1/2}$, where $\langle A,B\rangle
:=\mathrm{Tr}(A^{\dagger}B)$. Time reversibility of an approximate method that
approximates the exact flow $\Phi_{t}$ at discrete times $t=n\Delta t$ ($n$
integer) by an iterated map $\Phi_{\mathrm{appr},t}^{(\Delta t)}%
:=(\Phi_{\mathrm{appr}}^{(\Delta t)})^{n}:x_{\mathrm{eff},0}\mapsto
x_{\mathrm{eff},t}^{(\Delta t)}$ is measured by the distance
\begin{equation}
\mathcal{T}_{t}:=\Vert x_{\mathrm{eff},t}^{\mathrm{fb}}-x_{\mathrm{eff}%
,0}\Vert\label{eq:time_reversibility}%
\end{equation}
of the forward-backward propagated state $x_{\mathrm{eff},t}^{\mathrm{fb}%
}:=S\Phi_{\mathrm{appr},t}^{(\Delta t)}[S\Phi_{\mathrm{appr},t}^{(\Delta
t)}(x_{\mathrm{eff},0})]$ from the initial state $x_{\mathrm{eff},0}$. The
norm $\Vert x_{\mathrm{eff}}\Vert:=\langle x_{\mathrm{eff}},x_{\mathrm{eff}%
}\rangle^{1/2}$ of an effective phase space point $x_{\mathrm{eff}}$ is
defined using the scalar product $\langle x_{\mathrm{eff},1},x_{\mathrm{eff}%
,2}\rangle:=Q_{1}^{T} \cdot Q_{2}+P_{1}^{T} \cdot P_{2}+\langle\psi_{1}%
|\psi_{2}\rangle$ of $x_{\mathrm{eff},1}$ and $x_{\mathrm{eff},2}$. The
corresponding squared \textquotedblleft distance\textquotedblright\ $\Vert
x_{\mathrm{eff},1}-x_{\mathrm{eff},2}\Vert^{2}$ between points
$x_{\mathrm{eff},1}$ and $x_{\mathrm{eff},2}$ is simply the sum $\Vert
Q_{1}-Q_{2}\Vert^{2}+\Vert P_{1}-P_{2}\Vert^{2}+\Vert\psi_{1}-\psi_{2}%
\Vert^{2}$ of the squared distances between $Q_{1}$ and $Q_{2}$, between
$P_{1}$ and ${P}_{2}$, and between $\psi_{1}$ and $\psi_{2}$.

Panels~(g) and (h) of Fig.~\ref{fig:geom_prop} show that the energy is only
conserved approximately, to the same order as the order of convergence of the
integrator. The loss of exact energy conservation is standard for any
splitting method\cite{book_Hairer_Wanner:2006, Roulet_Vanicek:2019} and is due
to alternating kinetic and potential propagations: the effective Hamiltonian
alternates between $H_{\mathrm{eff}}=\langle\hat{T}_{\mathrm{el}%
}+T_{\mathrm{nu}}(P)\rangle_{\psi}$ and $H_{\mathrm{eff}}=\langle\hat
{V}(Q)\rangle_{\psi}$, and its time-dependent nature breaks the conservation
of energy.

\begin{figure}
\includegraphics[scale=1.0]{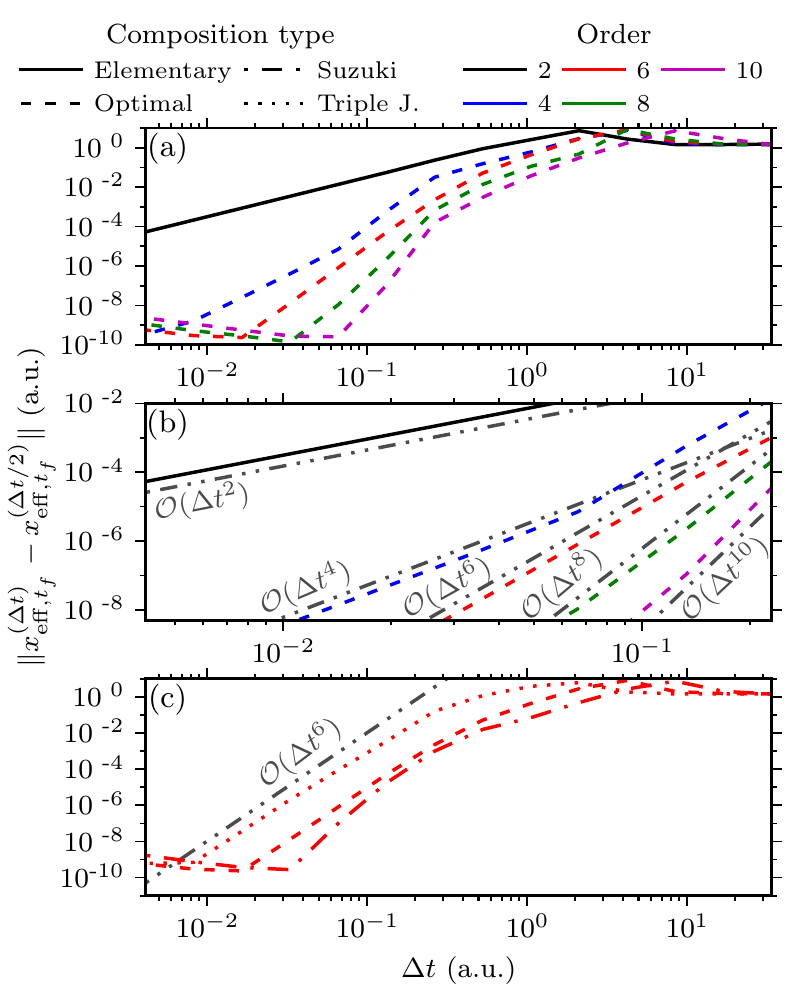}
\caption{Convergence of geometric integrators for Ehrenfest dynamics measured by the convergence error of the
final effective phase space point $x_{\mathrm{eff},t_{f}}$ as a function of $\Delta t$. Gray straight lines indicate
various predicted orders of convergence $\mathcal{O}(\Delta t^{n})$. (a) Methods obtained using the optimal
composition schemes, i.e., methods presented in Fig.~\ref{fig:geom_prop}. (b) Zoomed-in version of panel~(a), highlighting the asymptotic orders of convergence of the integrators. (c) Sixth-order methods obtained using
the triple jump, Suzuki-fractal, and optimal composition schemes. The error of an approximate method is measured
by the distance
$\| x_{\mathrm{eff},t_{f}}^{(\Delta t)} - x_{\mathrm{eff},t_{f}}^{(\Delta t/2)} \|$ of the final point
$ x_{\mathrm{eff},t_{f}}^{(\Delta t)} = \Phi_{\mathrm{appr},t_{f}}^{(\Delta t)}( x_{\mathrm{eff},0})$, obtained with
time step $\Delta t$, from the final point
$ x_{\mathrm{eff},t_{f}}^{(\Delta t/2)} = \Phi_{\mathrm{appr},t_{f}}^{(\Delta t/2)} (x_{\mathrm{eff},0})$, obtained
with half time step $\Delta t/2$.}\label{fig:convergence}
\end{figure}

Figure~\ref{fig:convergence} confirms the predicted asymptotic order of
convergence of the geometric integrators. However, panel~(c) may
wrongly suggest that the Suzuki-fractal scheme leads to the most efficient
method, as it has the smallest error for each time step size. What
Fig.~\ref{fig:convergence} does not show is that the sixth-order
Suzuki-fractal scheme has a factor of $25/9$ more substeps per time step than
either the triple-jump or optimal scheme. If we instead consider the
dependence of the convergence error on the computational cost [measured by the
central processing unit (CPU) time], the optimal composition scheme indeed
yields the most efficient method for each order of accuracy (see Fig.~S1 in
Sec.~S2 of the supplementary material).

\begin{figure}
\includegraphics[scale=1.0]{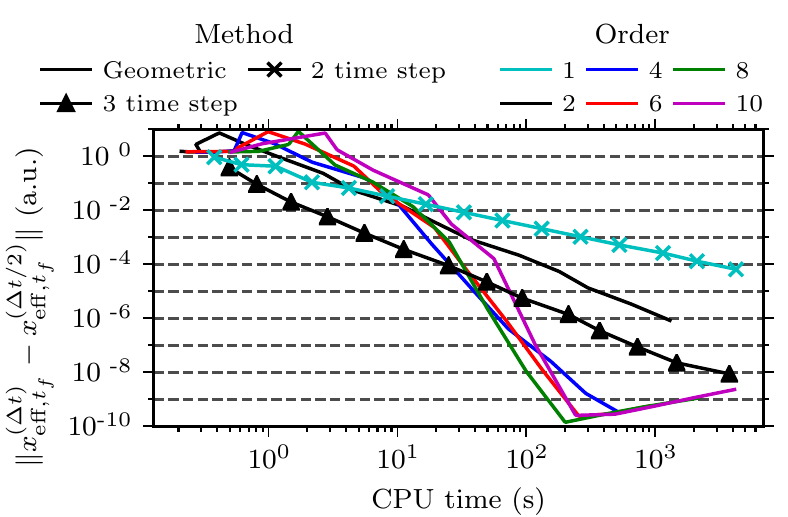}
\caption{Efficiency of the geometric integrators is compared with the efficiency of the widely-used two time step and three time step methods. Efficiency is measured using the dependence of the convergence error on the computational cost. Only the higher-order geometric integrators obtained using the optimal composition schemes are shown, for they are the most efficient for each order of accuracy.}\label{fig:efficiency}%

\end{figure}

To reach a modest convergence error of $10^{-3}$, the most efficient geometric
integrator (obtained using the optimal fourth-order composition scheme) is
$15$ times faster than the two time step method and roughly twice slower than
the three time step method (see Fig.~\ref{fig:efficiency}). Yet, a clear
advantage of the geometric integrators over the other methods is the exact
conservation of geometric properties. Both the two and three time step methods
violate symplecticity; the two time step method, in addition, violates time
reversibility (see Fig.~\ref{fig:geom_prop_vs_two_three_step} in
Appendix~\ref{sec:two_and_three_time_step}). Moreover, due to its higher order
of convergence in $\Delta t$, the fourth-order geometric integrator becomes
more efficient than even the three time step method to reach convergence
errors below $10^{-4}$ (see Fig.~\ref{fig:efficiency}).

\section{Conclusion \label{sec:conclusion}}

We have demonstrated that the high-order geometric integrators for Ehrenfest
dynamics can be obtained by simultaneously employing the splitting and
composition methods. Since Ehrenfest dynamics already involves a rather severe
approximation, one is often not interested in numerically converged solutions.
In such cases, geometric integrators become much more relevant because only
they guarantee the exact conservation of the geometric invariants regardless
of the accuracy of the solution.

That is not to say that the high-order geometric integrators are inefficient
for high accuracy simulations. On the contrary, to reach an error of, e.g.,
$10^{-6}$, using the eighth-order geometric integrator yields a four-fold
speedup over the three time step method (see Fig.~\ref{fig:efficiency}) and a
5000-fold speedup over the two time step method. High-accuracy results with
negligible numerical errors may be desirable when the error introduced by the
mean-field Ehrenfest approximation is either very small or unknown.

\section*{Supplementary material}

See the supplementary material for the computational details (Sec.~S1),
efficiency of the high-order geometric integrators obtained using the
triple-jump, Suzuki-fractal, optimal composition schemes (Sec.~S2), and
detailed algorithms of the two and three time step methods (Sec.~S3).

\section*{Acknowledgments}

The authors acknowledge the financial support from the European Research
Council (ERC) under the European Union's Horizon 2020 research and innovation
programme (grant agreement No. 683069 -- MOLEQULE) and thank Tomislav
Begu\v{s}i\'{c} and Nikolay Golubev for useful discussions.

\section*{Author Declarations}
\subsection*{Conflict of interest}
The authors have no conflicts to disclose.

\section*{Data Availability}

The data that support the findings of this study are openly
available in Zenodo at https://doi.org/10.5281/zenodo.5167211.

\begin{appendix}

\section{\label{sec:two_and_three_time_step} Two and three time
step methods}

\subsection{\label{sec:two_time_step} Two time step method}

Unlike the geometric integrators, which propagate $Q_{t},P_{t}$, and $\psi
_{t}$ simultaneously, the two time step method\cite{Feng_Runge:1991,
Micha_Runge:1994, Micha:1999} consists in alternately propagating the
classical nuclear phase space point and electronic wavefunction. The time step
$\Delta t_{\mathrm{el}}=\Delta t/n_{\mathrm{el}}$ for the electronic
propagation is typically much smaller than the time step $\Delta t$ for the
nuclear propagation (we used $n_{\mathrm{el}}=100$).

In the two time step method, we employed the second-order Verlet
algorithm\cite{Verlet:1967} to propagate the nuclear phase space point and the
second-order VTV split-operator algorithm\cite{Feit_Steiger:1982} to propagate
the electronic wavefunction. However, because the nuclear phase space point
and electronic wavefunction are propagated separately and alternately, the
overall two time step method is only first-order accurate in the time step.
Moreover, the method is neither time-reversible nor symplectic (see
Fig.~\ref{fig:geom_prop_vs_two_three_step}). See Sec.~S3 of the
supplementary material for the detailed algorithm of the two time step
method.

\subsection{\label{sec:three_time_step} Three time step method}

\begin{figure}
[ht!]\includegraphics[scale=1.0]{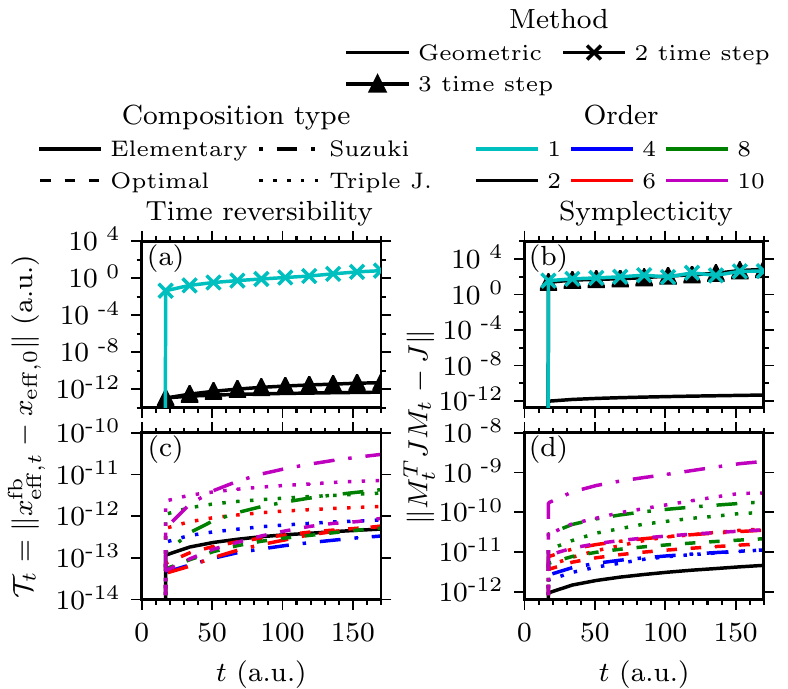}
\caption{Violation of (a) time reversibility [see Eq.~(\ref{eq:time_reversibility})] and (b) symplecticity [see Eq.~(\ref{eq:symplecticity})] by the non-geometric integrators: The two time step method is neither reversible nor symplectic, whereas the three
time step method is time-reversible but not symplectic. The geometric integrators exactly preserve both (c) time-reversibility and (d) symplecticity. Time step $\Delta t = 17$ a.u. was used. For the two and three time step methods, the corresponding convergence errors are $0.5$ and $0.09$, respectively, and for all of the presented geometric integrators, the errors are $> 1.4$.}\label{fig:geom_prop_vs_two_three_step}%

\end{figure}

The three time step method,\cite{Li_Frisch:2005, Ding_Li:2015} owing to its
symmetry, is both time-reversible and second-order accurate in the time
step.\cite{book_Hairer_Wanner:2006} However, the method is still not
symplectic (see Fig.~\ref{fig:geom_prop_vs_two_three_step}).

In addition to the nuclear time step $\Delta t$ and electronic time step
$\Delta t_{\mathrm{el}}$, used also in the two time step method, the three
time step method improves the efficiency by introducing the intermediate
\emph{nuclear-electronic coupling} time step $\Delta t_{\text{nu-el}} = \Delta
t / n_{\text{nu-el}} = \Delta t_{\mathrm{el}} n_{\mathrm{el}}/n_{\text{nu-el}%
}$ (we used $n_{\text{nu-el}}=10$). See Sec.~S3 of the
supplementary material for the detailed algorithm of the three time step
method.

\section{\label{sec:qm_symplec} Quantum canonical two-form}

The canonical two-form $dq_{\psi}\wedge dp_{\psi}$ acts on states $\psi_{1}$
and $\psi_{2}$ as%
\begin{equation}
dq_{\psi}\wedge dp_{\psi}(\psi_{1},\psi_{2})=2\hbar\mathrm{Im}\langle\psi
_{1}|\psi_{2}\rangle
\end{equation}
because
\begin{align}
dq_{\psi}  &  \wedge dp_{\psi}(\psi_{1},\psi_{2})\nonumber\\
&  =\langle dq_{\psi}(\psi_{1})|dp_{\psi}(\psi_{2})\rangle-\langle dp_{\psi
}(\psi_{1})|dq_{\psi}(\psi_{2})\rangle\nonumber\\
&  ={2\hbar}[\langle\mathrm{Re}\psi_{1}(q)|\mathrm{Im}\psi_{2}(q)\rangle
-\langle\mathrm{Im}\psi_{1}(q)|\mathrm{Re}\psi_{2}(q)\rangle]\nonumber\\
&  =2\hbar\mathrm{Im}\langle\psi_{1}|\psi_{2}\rangle,
\end{align}
where we have used that the tangent space of a vector space can be identified
with the vector space itself,\cite{book_Lee:2009} i.e.,
\begin{align}
dq_{\psi}(\psi)  &  =q_{\psi}=\sqrt{2\hbar}\mathrm{Re}\psi(q),\\
dp_{\psi}(\psi)  &  =p_{\psi}=\sqrt{2\hbar}\mathrm{Im}\psi(q).
\end{align}

\section{\label{sec:Ham_flow_symplec} Symplecticity of the exact
Hamiltonian flow}

We use the \emph{standard symplectic matrix}
\begin{equation}
J:=%
\begin{pmatrix}
0 & -I_{D+N}\\
I_{D+N} & 0
\end{pmatrix}
\end{equation}
and re-express effective symplectic two-form~(\ref{eq:mqc_two_form})
as\cite{book_Leimkuhler_Reich:2004}
\begin{equation}
\omega_{\mathrm{eff}}=\frac{1}{2}(Jdx_{\mathrm{eff}})\wedge dx_{\mathrm{eff}}.
\end{equation}
Since $dx_{\mathrm{eff},t}=M_{t}dx_{\mathrm{eff},0}$, we have
\begin{align}
(Jdx_{\mathrm{eff},t})\wedge dx_{\mathrm{eff},t}  &  =(JM_{t}dx_{\mathrm{eff}%
,0})\wedge(M_{t}dx_{\mathrm{eff},0})\nonumber\\
&  =(M_{t}^{T}JM_{t}dx_{\mathrm{eff},0})\wedge dx_{\mathrm{eff},0},
\end{align}
and two-form $\omega_{\mathrm{eff}}$ is conserved [i.e., $(Jdx_{\mathrm{eff}%
,t})\wedge dx_{\mathrm{eff},t}=(Jdx_{\mathrm{eff},0})\wedge dx_{\mathrm{eff}%
,0}$] if $M_{t}$ is a symplectic matrix, i.e., if it satisfies the condition
\begin{equation}
M_{t}^{T}JM_{t}=J. \label{eq:symplecticity_condition}%
\end{equation}

Since $M_{0}=I$, Eq.~(\ref{eq:symplecticity_condition}) is trivially satisfied
at $t=0$. To show that the stability matrix $M_{t}$ of Hamiltonian flow
$\Phi_{H_{\mathrm{eff}},t}$ is symplectic, we therefore only have to show that
$d(M_{t}^{T}JM_{t})/dt=0$, which follows easily from the calculation
\begin{align}
\frac{d}{dt}M_{t}^{T}JM_{t}  &  =\dot{M}_{t}^{T}JM_{t}+M_{t}^{T}J\dot{M}%
_{t}\nonumber\\
&  =M_{t}^{T}\mathrm{Hess}[H_{\mathrm{eff}}(x_{\mathrm{eff},t})]J^{2}
M_{t}\nonumber\\
&  \qquad\qquad+ M_{t}^{T}J J^{T}\mathrm{Hess}[H_{\mathrm{eff}}%
(x_{\mathrm{eff},t})]M_{t}\nonumber\\
&  = - M_{t}^{T}\mathrm{Hess}[H_{\mathrm{eff}}(x_{\mathrm{eff},t}%
)]M_{t}\nonumber\\
&  \qquad\qquad+ M_{t}^{T}\mathrm{Hess}[H_{\mathrm{eff}}(x_{\mathrm{eff}%
,t})]M_{t}\nonumber\\
&  =0, \label{eq:symplecticity_proof}%
\end{align}
where we have used the fact that the time derivative of the stability matrix
satisfies\cite{book_Leimkuhler_Reich:2004}
\begin{equation}
\dot{M_{t}}=J^{T} \mathrm{Hess}[H_{\mathrm{eff}}(x_{\mathrm{eff},t})]M_{t}.
\label{eq:stability_matrix_t_dep}%
\end{equation}
Although we did not need the explicit form 
\begin{align}
&  \mathrm{Hess}[H_{\mathrm{eff}}(x_{\mathrm{eff}})] =\nonumber\\
&
\begin{pmatrix}
\langle\mathrm{Hess}\hat{V}(Q)\rangle_{\psi} & d_{V}(Q)q_{\psi} & 0 &
d_{V}(Q)p_{\psi}\\
d_{V}(Q) q_{\psi} & \hat{H}_{\mathrm{el}}(Q,P)/\hbar & d_{T}(P) q_{\psi} & 0\\
0 & d_{T}(P)q_{\psi} & 1/M & d_{T}(P)p_{\psi}\\
d_{V}(Q) p_{\psi} & 0 & d_{T}(P)p_{\psi} & \hat{H}_{\mathrm{el}}(Q,P)/\hbar
\end{pmatrix}
\label{eq:Hess_H_eff}%
\end{align}
of the Hessian of $H_{\mathrm{eff}}(x_{\mathrm{eff}})$ to prove the
symplecticity of $M_{t}$, this expression will be useful for the numerical
propagation of the stability matrix; we defined and used
\begin{align}
d_{V}(Q)  &  := \hat{V}^{\prime}(Q)/\hbar,\\
d_{T}(P)  &  := P/(\hbar M)
\end{align}
to simplify Eq.~(\ref{eq:Hess_H_eff}). 

\section{\label{sec:Num_prop_stab_mat} Numerical propagation of
the stability matrix}

The numerical propagation of stability matrix $M_{t}$ requires much more
computational effort than the propagation of mixed quantum-classical phase
space point $x_{\mathrm{eff}}$. It is, in general, not necessary to propagate
the stability matrix to simulate Ehrenfest dynamics. Yet, to numerically
demonstrate the symplecticity of the geometric integrators, i.e., to prepare
panels~(e) and (f) of Fig.~\ref{fig:geom_prop}, we propagated also the
stability matrix. Like in the propagation of $x_{\mathrm{eff}}$, we only need
to present the analytical solutions of the kinetic and potential propagation
steps for arbitrary times $t$ because all presented geometric integrators are
composed of kinetic and potential propagations (see
Sec.~\ref{subsec:geometric_integrators}).

During the kinetic propagation, the equation of motion
\begin{equation}
\dot{M_{t}}=J^{T}\mathrm{Hess}[\langle\hat{T}_{\mathrm{nu}+\mathrm{el}}%
(P_{t})\rangle_{\psi_{t}}] M_{t} \label{eq:stability_matrix_T_prop}%
\end{equation}
for stability matrix $M_{t}$ is obtained by reducing the effective Hamiltonian
to $H_{\mathrm{eff}}(x_{\mathrm{eff}}) =\langle\hat{T}_{\mathrm{nu}%
+\mathrm{el}}(P)\rangle_{\psi}=\langle T_{\mathrm{nu}}(P) + \hat
{T}_{\mathrm{el}}\rangle_{\psi}$ in Eq.~(\ref{eq:stability_matrix_t_dep}). The
explicit form of the Hessian is
\begin{align}
\mathrm{Hess}  &  [\langle\hat{T}_{\mathrm{nu}+\mathrm{el}}(P)\rangle_{\psi}]
=\nonumber\\
&
\begin{pmatrix}
0 & 0 & 0 & 0\\
0 & \hat{T}_{\mathrm{nu}+\mathrm{el}}(P)/\hbar & d_{T}(P) q_{\psi} & 0\\
0 & d_{T}(P)q_{\psi} & 1/M & d_{T}(P)p_{\psi}\\
0 & 0 & d_{T}(P)p_{\psi} & \hat{T}_{\mathrm{nu}+\mathrm{el}}(P)/\hbar
\end{pmatrix}
,
\end{align}
and Eq.~(\ref{eq:stability_matrix_T_prop}) can be solved analytically to
yield
\begin{equation}
M_{t}=
\begin{pmatrix}
1 & td_{T}(P_{0})q_{\psi,0} & t/M & td_{T}(P_{0})p_{\psi,0}\\
0 & c_{T}(P_{0}) & td_{T}(P_{0})p_{\psi,t} & s_{T}(P_{0})\\
0 & 0 & 1 & 0\\
0 & -s_{T}(P_{0}) & -t d_{T}(P_{0})q_{\psi,t} & c_{T}(P_{0})
\end{pmatrix}
M_{0},
\end{equation}
where
\begin{align}
c_{T}(P) := \cos[\hat{T}_{\mathrm{nu}+\mathrm{el}}(P)t/\hbar],\\
s_{T}(P) := \sin[\hat{T}_{\mathrm{nu}+\mathrm{el}}(P)t/\hbar],
\end{align}
and the propagation of quantum Darboux coordinates
\begin{align}
q_{\psi,t}  &  =c_{T}(P_{0}) q_{\psi,0}+s_{T}(P_{0})p_{\psi,0},\\
p_{\psi,t}  &  =-s_{T}(P_{0})q_{\psi,0}+c_{T}(P_{0})p_{\psi,0}%
\end{align}
for time $t$ is equivalent to the standard propagation of electronic
wavefunction $\psi_{t}=\exp[-it\hat{T}_{\mathrm{nu}+\mathrm{el}}(P_{0}%
)/\hbar]\psi_{0}$ since $\psi_{t}=(q_{\psi,t}+ip_{\psi,t})/\sqrt{2\hbar}$.

During the potential propagation, the equation of motion
\begin{equation}
\dot{M_{t}}=J^{T} \mathrm{Hess}[\langle\hat{V}(Q_{t}) \rangle_{\psi_{t}}]
M_{t} \label{eq:stability_matrix_V_prop}%
\end{equation}
for stability matrix $M_{t}$ is obtained by reducing the effective Hamiltonian
to $H_{\mathrm{eff}} (x_{\mathrm{eff}}) = \langle\hat{V}(Q) \rangle_{\psi}$ in
Eq.~(\ref{eq:stability_matrix_t_dep}). The explicit form of the Hessian is
\begin{align}
\mathrm{Hess}  &  [\langle\hat{V}(Q) \rangle_{\psi}] =\nonumber\\
&
\begin{pmatrix}
\langle\mathrm{Hess}\hat{V}(Q)\rangle_{\psi} & d_{V}(Q)q_{\psi} & 0 &
d_{V}(Q)p_{\psi}\\
d_{V}(Q)q_{\psi} & \hat{V}(Q)/\hbar & 0 & 0\\
0 & 0 & 0 & 0\\
d_{V}(Q)p_{\psi} & 0 & 0 & \hat{V}(Q)/\hbar
\end{pmatrix}
,
\end{align}
and Eq.~(\ref{eq:stability_matrix_V_prop}) can be solved analytically to
yield
\begin{align}
&  M_{t}=\nonumber\\
&
\begin{pmatrix}
1 & 0 & 0 & 0\\
t d_{V}(Q_{0}) p_{\psi,t} & c_{V}(Q_{0}) & 0 & s_{V}(Q_{0})\\
-t\langle\mathrm{Hess}\hat{V}(Q_{0})\rangle_{\psi_{0}} & -t d_{V}%
(Q_{0})q_{\psi,0} & 1 & -t d_{V}(Q_{0})p_{\psi,0}\\
-t d_{V}(Q_{0})q_{\psi,t} & -s_{V}(Q_{0}) & 0 & c_{V}(Q_{0})
\end{pmatrix}
M_{0},
\end{align}
where
\begin{align}
c_{V}(Q) := \cos[\hat{V}(Q)t/\hbar],\\
s_{V}(Q) := \sin[\hat{V}(Q)t/\hbar],
\end{align}
and the propagation of quantum Darboux coordinates
\begin{align}
q_{\psi,t}  &  =c_{V}(Q_{0})q_{\psi,0}+s_{V}(Q_{0})p_{\psi,0},\\
p_{\psi,t}  &  =-s_{V}(Q_{0})q_{\psi,0}+c_{V}(Q_{0})p_{\psi,0}%
\end{align}
for time $t$ is equivalent to the propagation of the electronic wavefunction
$\psi_{t}=\exp[-it\hat{V}(Q_{0})/\hbar]\psi_{0}$ since $\psi_{t}=(q_{\psi
,t}+ip_{\psi,t})/\sqrt{2\hbar}$.
\end{appendix}

\setcounter{section}{0}
\setcounter{equation}{0}
\setcounter{figure}{0}
\setcounter{table}{0}
\renewcommand{\theequation}{S\arabic{equation}}
\renewcommand{\thefigure}{S\arabic{figure}} \renewcommand{\thesection}{S\arabic{section}}
\clearpage
\textbf{\Large{Supplementary material for: \\ High-order geometric integrators for representation-free Ehrenfest dynamics}} 

\vspace{5.0ex}
This document provides information supporting the main text. It contains the
computational details (Sec.~S1), efficiency of the high-order
geometric integrators obtained using the triple-jump, Suzuki-fractal, and
optimal composition schemes (Sec.~S2), and detailed algorithms of the two and three time step methods (Sec.~S3).
\vspace{7.0ex}
\section{\label{sec:Computational_details} Computational details}

In the exact quantum simulation, the full wavefunction $\Psi_{t}$ was
represented on a uniform four-dimensional grid that is a tensor product of two
different two-dimensional grids for the nuclear and electronic degrees of
freedom. A uniform grid of $101\times101$ points defined between $Q_{n}=-3$
a.u. and $Q_{n} =3$ a.u. was used for the nuclear degrees of freedom. As for
the electronic degrees of freedom, a uniform grid of $64 \times64$ points
defined between $q_{n} = -10$ a.u. and $q_{n} = 10$ a.u. was used; this grid
was also used to represent the electronic wavefunction in Ehrenfest
simulations. We employed the second-order VTV split-operator algorithm with
time step $\Delta t = 0.1$ a.u. for the numerical propagation of $\Psi_{t}$.

To obtain the initial states, we solved the electronic time-independent
Schr\"{o}dinger equation [Eq.~(53) of the main text] with the imaginary time
propagation method:\cite{Kosloff_Tal-Ezer:1986} To prepare quantum initial
state~(51) of the main text, we solved the equation at every point on the
nuclear grid. In contrast, to obtain the initial effective phase space point
for Ehrenfest simulations [see Eq.~(55) of the main text], we only had to
solve the equation once at the initial nuclear position $Q_{0}$.
\clearpage

\section{\label{sec:geom_int_efficiency} Efficiency of the geometric
integrators}

\begin{figure}
[h!]\includegraphics[scale=1.0]{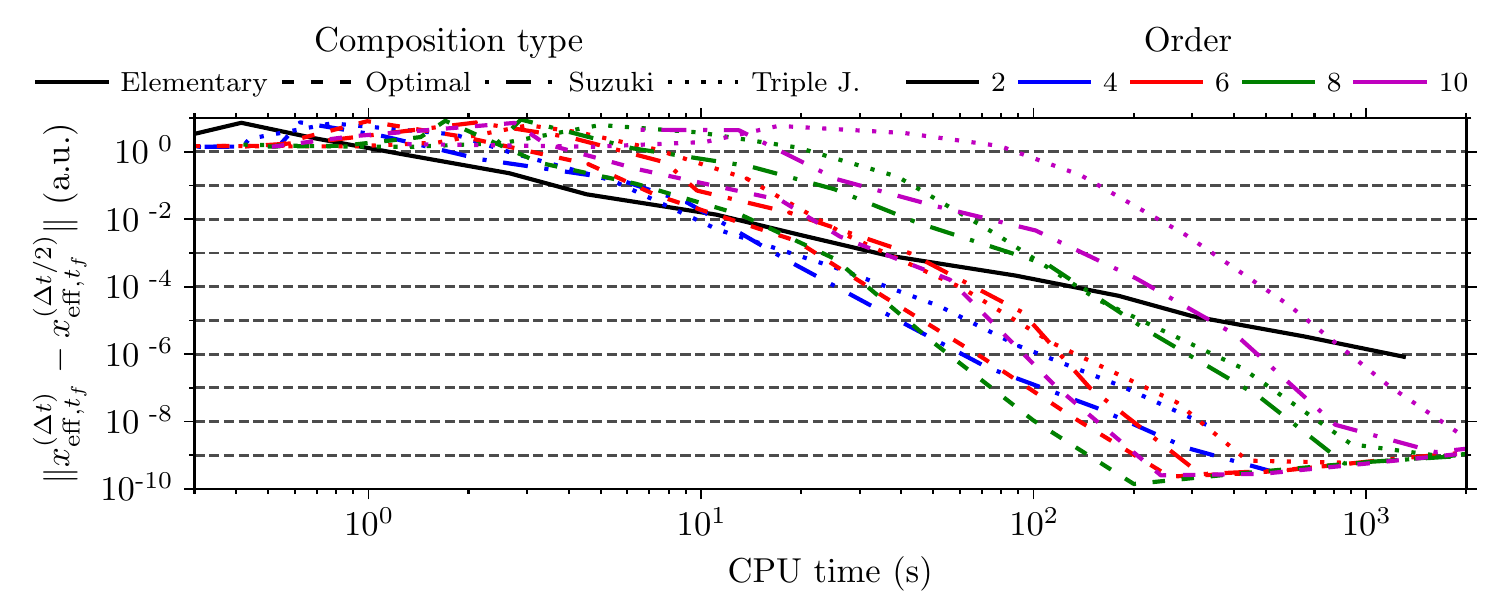}
\caption{Efficiency of the geometric integrators presented in Sec.~II\,D of the main text.
Like in the main text, efficiency is measured using the dependence of the convergence error on the computational
cost (which we measure by CPU time).
As expected, for each order of accuracy, the optimal composition scheme yields the most efficient method.}\label{fig:efficiency_all_methods}%

\end{figure}

\section{Detailed algorithms of the two and three time step methods}

\subsection{Two time step method}

The detailed algorithm of the two time step method is as follows (like in the main text, $\Delta t_{\mathrm{el}} = \Delta t/n_{\mathrm{el}}$):\newline%
(\texttt{{\small {Verlet algorithm with time step $\Delta t$}}}:)
\newline$P_{t+\Delta t/2}=P_{t}-\frac{\Delta t}{2}\langle\hat{V}^{\prime
}(Q_{t})\rangle_{\psi_{t}}$ \newline$Q_{t+\Delta t}=Q_{t}+\Delta tM^{-1}\cdot
P_{t+\Delta t/2}$ \newline$P_{t+\Delta t}=P_{t+\Delta t/2}-\frac{\Delta t}%
{2}\langle\hat{V}^{\prime}(Q_{t+\Delta t})\rangle_{\psi_{t}}$ \newline%
\texttt{Do} $i=1,\dots,n_{\mathrm{el}}$ \newline\indent(\texttt{{\small {VTV
split-operator algorithm with time step $\Delta t_{\mathrm{el}}$}}}:)
\newline\indent$\psi_{t+i\Delta t_{\mathrm{el}}}=\hat{U}_{\mathrm{VTV}%
}^{(\Delta t_{\mathrm{el}})}(Q_{t+\Delta t},P_{t+\Delta t})\psi_{t+(i-1)\Delta
t_{\mathrm{el}}}$ \newline\texttt{End Do,}\newline where the evolution
operator
\begin{equation}
\hat{U}_{\mathrm{VTV}}^{(\tau)}(Q,P)=\hat{U}_{\mathrm{V}}^{(\tau/2)}(Q)\hat
{U}_{\mathrm{T}}^{(\tau)}(P)\hat{U}_{\mathrm{V}}^{(\tau/2)}(Q)
\end{equation}
for the VTV split-operator algorithm is composed of the potential and kinetic
propagation steps,
\begin{align}
\hat{U}_{\mathrm{V}}^{(\tau)}(Q)  &  =\exp{[-i\tau\hat{V}(Q)/\hbar]},\\
\hat{U}_{\mathrm{T}}^{(\tau)}(P)  &  =\exp{[-i\tau\hat{T}_{\mathrm{nu}%
+\mathrm{el}}(P)/\hbar]};
\end{align}
here, $\hat{T}_{\mathrm{nu}+\mathrm{el}}(P)=T_{\mathrm{nu}}(P)+\hat
{T}_{\mathrm{el}}$.

\subsection{Three time step method}

The detailed algorithm of the three time step method is as follows (like in the main text, $\Delta t_{\textrm{nu-el}} = \Delta t/n_{\textrm{nu-el}} = \Delta t_{\mathrm{el}} n_{\mathrm{el}} / n_{\textrm{nu-el}}$):\newline%
(\texttt{{\small {Propagation of nuclear momentum by $\Delta t/2$}}}:)
\newline$P_{t+\Delta t/2}=P_{t}-\frac{\Delta t}{2}\langle\hat{V}^{\prime
}(Q_{t})\rangle_{\psi_{t}}$ \newline\texttt{Do} $i=1,\dots,n_{\text{nu-el}}$
\newline\indent(\texttt{{\small {Propagation of nuclear position by $\Delta
t_{\text{nu-el}}/2$}}}:) \newline\indent$Q_{t+(2i-1)\Delta t_{\text{nu-el}}%
/2}=Q_{t+(i-1)\Delta t_{\text{nu-el}}}+\frac{\Delta t_{\text{nu-el}}}{2}%
M^{-1}\cdot P_{t+\Delta t/2}$\newline\indent\texttt{Do} $j=1,\dots
,n_{\mathrm{el}}/n_{\text{nu-el}}$ \newline\indent\indent(\texttt{{\small {VTV
split-operator algorithm with the time step of $\Delta t_{\mathrm{el}}$}}}:)
\newline\indent\indent$\psi_{t+(i-1)\Delta t_{\text{nu-el}}+j\Delta
t_{\mathrm{el}}}=\hat{U}_{\mathrm{VTV}}^{(\Delta t_{\mathrm{el}}%
)}(Q_{t+(2i-1)\Delta t_{\text{nu-el}}/2},P_{t+\Delta t/2})\psi_{t+(i-1)\Delta
t_{\text{nu-el}}+(j-1)\Delta t_{\mathrm{el}}}$ \newline\indent\texttt{End
Do}\newline\indent(\texttt{{\small {Propagation of nuclear position by $\Delta
t_{\text{nu-el}}/2$}}}:) \newline\indent$Q_{t+i\Delta t_{\text{nu-el}}%
}=Q_{t+(2i-1)\Delta t_{\text{nu-el}}/2}+\frac{\Delta t_{\text{nu-el}}}%
{2}M^{-1}\cdot P_{t+\Delta t/2}$\newline\texttt{End Do}\newline%
(\texttt{{\small {Propagation of nuclear momentum by $\Delta t/2$}}}:)
\newline$P_{t+\Delta t}=P_{t+\Delta t/2}-\frac{\Delta t}{2}\langle\hat
{V}^{\prime}(Q_{t+\Delta t})\rangle_{\psi_{t+\Delta t}}$. \newline
\clearpage

\bibliographystyle{aipnum4-2}
\bibliography{Ehrenfest_rep_independent}

%aipnum4-2.bst 2019-01-14 (MD) hand-edited version of apsrev4-1.bst
%Control: key (0)
%Control: author (8) initials jnrlst
%Control: editor formatted (1) identically to author
%Control: production of article title (-1) disabled
%Control: page (0) single
%Control: year (1) truncated
%Control: production of eprint (0) enabled
\begin{thebibliography}{113}%
\makeatletter
\providecommand \@ifxundefined [1]{%
 \@ifx{#1\undefined}
}%
\providecommand \@ifnum [1]{%
 \ifnum #1\expandafter \@firstoftwo
 \else \expandafter \@secondoftwo
 \fi
}%
\providecommand \@ifx [1]{%
 \ifx #1\expandafter \@firstoftwo
 \else \expandafter \@secondoftwo
 \fi
}%
\providecommand \natexlab [1]{#1}%
\providecommand \enquote  [1]{``#1''}%
\providecommand \bibnamefont  [1]{#1}%
\providecommand \bibfnamefont [1]{#1}%
\providecommand \citenamefont [1]{#1}%
\providecommand \href@noop [0]{\@secondoftwo}%
\providecommand \href [0]{\begingroup \@sanitize@url \@href}%
\providecommand \@href[1]{\@@startlink{#1}\@@href}%
\providecommand \@@href[1]{\endgroup#1\@@endlink}%
\providecommand \@sanitize@url [0]{\catcode `\\12\catcode `\$12\catcode
  `\&12\catcode `\#12\catcode `\^12\catcode `\_12\catcode `\%12\relax}%
\providecommand \@@startlink[1]{}%
\providecommand \@@endlink[0]{}%
\providecommand \url  [0]{\begingroup\@sanitize@url \@url }%
\providecommand \@url [1]{\endgroup\@href {#1}{\urlprefix }}%
\providecommand \urlprefix  [0]{URL }%
\providecommand \Eprint [0]{\href }%
\providecommand \doibase [0]{https://doi.org/}%
\providecommand \selectlanguage [0]{\@gobble}%
\providecommand \bibinfo  [0]{\@secondoftwo}%
\providecommand \bibfield  [0]{\@secondoftwo}%
\providecommand \translation [1]{[#1]}%
\providecommand \BibitemOpen [0]{}%
\providecommand \bibitemStop [0]{}%
\providecommand \bibitemNoStop [0]{.\EOS\space}%
\providecommand \EOS [0]{\spacefactor3000\relax}%
\providecommand \BibitemShut  [1]{\csname bibitem#1\endcsname}%
\let\auto@bib@innerbib\@empty
%</preamble>
\bibitem [{\citenamefont {Tully}\ and\ \citenamefont
  {Preston}(1971)}]{Tully_Preston:1971}%
  \BibitemOpen
  \bibfield  {author} {\bibinfo {author} {\bibfnamefont {J.~C.}\ \bibnamefont
  {Tully}}\ and\ \bibinfo {author} {\bibfnamefont {R.~K.}\ \bibnamefont
  {Preston}},\ }\href@noop {} {\bibfield  {journal} {\bibinfo  {journal}
  {J.~Chem.\ Phys.}\ }\textbf {\bibinfo {volume} {55}},\ \bibinfo {pages} {562}
  (\bibinfo {year} {1971})}\BibitemShut {NoStop}%
\bibitem [{\citenamefont {Tully}(1990)}]{Tully:1990}%
  \BibitemOpen
  \bibfield  {author} {\bibinfo {author} {\bibfnamefont {J.~C.}\ \bibnamefont
  {Tully}},\ }\href {https://doi.org/10.1063/1.459170} {\bibfield  {journal}
  {\bibinfo  {journal} {J.~Chem.\ Phys.}\ }\textbf {\bibinfo {volume} {93}},\
  \bibinfo {pages} {1061} (\bibinfo {year} {1990})}\BibitemShut {NoStop}%
\bibitem [{\citenamefont {Schmidt}, \citenamefont {Parandekar},\ and\
  \citenamefont {Tully}(2008)}]{Schmidt_Tully:2008}%
  \BibitemOpen
  \bibfield  {author} {\bibinfo {author} {\bibfnamefont {J.~R.}\ \bibnamefont
  {Schmidt}}, \bibinfo {author} {\bibfnamefont {P.~V.}\ \bibnamefont
  {Parandekar}},\ and\ \bibinfo {author} {\bibfnamefont {J.~C.}\ \bibnamefont
  {Tully}},\ }\href {https://doi.org/10.1063/1.2955564} {\bibfield  {journal}
  {\bibinfo  {journal} {J.~Chem.\ Phys.}\ }\textbf {\bibinfo {volume} {129}},\
  \bibinfo {pages} {044104} (\bibinfo {year} {2008})}\BibitemShut {NoStop}%
\bibitem [{\citenamefont {Lasser}\ and\ \citenamefont
  {Swart}(2008)}]{Lasser_Swart:2008}%
  \BibitemOpen
  \bibfield  {author} {\bibinfo {author} {\bibfnamefont {C.}~\bibnamefont
  {Lasser}}\ and\ \bibinfo {author} {\bibfnamefont {T.}~\bibnamefont {Swart}},\
  }\href {https://doi.org/10.1063/1.2954019} {\bibfield  {journal} {\bibinfo
  {journal} {J.~Chem.\ Phys.}\ }\textbf {\bibinfo {volume} {129}},\ \bibinfo
  {pages} {034302} (\bibinfo {year} {2008})}\BibitemShut {NoStop}%
\bibitem [{\citenamefont {Subotnik}\ and\ \citenamefont
  {Shenvi}(2011)}]{Subotnik_Shenvi:2011}%
  \BibitemOpen
  \bibfield  {author} {\bibinfo {author} {\bibfnamefont {J.~E.}\ \bibnamefont
  {Subotnik}}\ and\ \bibinfo {author} {\bibfnamefont {N.}~\bibnamefont
  {Shenvi}},\ }\href {https://doi.org/10.1063/1.3506779} {\bibfield  {journal}
  {\bibinfo  {journal} {J.~Chem.\ Phys.}\ }\textbf {\bibinfo {volume} {134}},\
  \bibinfo {pages} {024105} (\bibinfo {year} {2011})}\BibitemShut {NoStop}%
\bibitem [{\citenamefont {Billing}(1975)}]{Billing:1975}%
  \BibitemOpen
  \bibfield  {author} {\bibinfo {author} {\bibfnamefont {G.~D.}\ \bibnamefont
  {Billing}},\ }\href
  {https://doi.org/https://doi.org/10.1016/0009-2614(75)80014-5} {\bibfield
  {journal} {\bibinfo  {journal} {Chem.\ Phys.\ Lett.}\ }\textbf {\bibinfo
  {volume} {30}},\ \bibinfo {pages} {391} (\bibinfo {year} {1975})}\BibitemShut
  {NoStop}%
\bibitem [{\citenamefont {Billing}(1976)}]{Billing:1976}%
  \BibitemOpen
  \bibfield  {author} {\bibinfo {author} {\bibfnamefont {G.~D.}\ \bibnamefont
  {Billing}},\ }\href {https://doi.org/10.1063/1.432205} {\bibfield  {journal}
  {\bibinfo  {journal} {J.~Chem.\ Phys.}\ }\textbf {\bibinfo {volume} {64}},\
  \bibinfo {pages} {908} (\bibinfo {year} {1976})}\BibitemShut {NoStop}%
\bibitem [{\citenamefont {Tully}(1998)}]{Tully:1998}%
  \BibitemOpen
  \bibfield  {author} {\bibinfo {author} {\bibfnamefont {J.~C.}\ \bibnamefont
  {Tully}},\ }\href@noop {} {\bibfield  {journal} {\bibinfo  {journal} {Faraday
  Discuss.}\ }\textbf {\bibinfo {volume} {110}},\ \bibinfo {pages} {407}
  (\bibinfo {year} {1998})}\BibitemShut {NoStop}%
\bibitem [{\citenamefont {Micha}(1983)}]{Micha:1983}%
  \BibitemOpen
  \bibfield  {author} {\bibinfo {author} {\bibfnamefont {D.~A.}\ \bibnamefont
  {Micha}},\ }\href {https://doi.org/10.1063/1.444753} {\bibfield  {journal}
  {\bibinfo  {journal} {J.~Chem.\ Phys.}\ }\textbf {\bibinfo {volume} {78}},\
  \bibinfo {pages} {7138} (\bibinfo {year} {1983})}\BibitemShut {NoStop}%
\bibitem [{\citenamefont {Sawada}, \citenamefont {Nitzan},\ and\ \citenamefont
  {Metiu}(1985)}]{Sawada_Metiu:1985}%
  \BibitemOpen
  \bibfield  {author} {\bibinfo {author} {\bibfnamefont {S.-I.}\ \bibnamefont
  {Sawada}}, \bibinfo {author} {\bibfnamefont {A.}~\bibnamefont {Nitzan}},\
  and\ \bibinfo {author} {\bibfnamefont {H.}~\bibnamefont {Metiu}},\ }\href
  {https://doi.org/10.1103/PhysRevB.32.851} {\bibfield  {journal} {\bibinfo
  {journal} {Phys.\ Rev.~B}\ }\textbf {\bibinfo {volume} {32}},\ \bibinfo
  {pages} {851} (\bibinfo {year} {1985})}\BibitemShut {NoStop}%
\bibitem [{\citenamefont {Micha}\ and\ \citenamefont
  {Runge}(1994)}]{Micha_Runge:1994}%
  \BibitemOpen
  \bibfield  {author} {\bibinfo {author} {\bibfnamefont {D.~A.}\ \bibnamefont
  {Micha}}\ and\ \bibinfo {author} {\bibfnamefont {K.}~\bibnamefont {Runge}},\
  }\href {https://doi.org/10.1103/PhysRevA.50.322} {\bibfield  {journal}
  {\bibinfo  {journal} {Phys.\ Rev.~A}\ }\textbf {\bibinfo {volume} {50}},\
  \bibinfo {pages} {322} (\bibinfo {year} {1994})}\BibitemShut {NoStop}%
\bibitem [{\citenamefont {Micha}(1999)}]{Micha:1999}%
  \BibitemOpen
  \bibfield  {author} {\bibinfo {author} {\bibfnamefont {D.~A.}\ \bibnamefont
  {Micha}},\ }\href {https://doi.org/10.1021/jp9906839} {\bibfield  {journal}
  {\bibinfo  {journal} {J.~Phys.\ Chem.~A}\ }\textbf {\bibinfo {volume}
  {103}},\ \bibinfo {pages} {7562} (\bibinfo {year} {1999})}\BibitemShut
  {NoStop}%
\bibitem [{\citenamefont {Li}\ \emph {et~al.}(2005{\natexlab{a}})\citenamefont
  {Li}, \citenamefont {Tully}, \citenamefont {Schlegel},\ and\ \citenamefont
  {Frisch}}]{Li_Frisch:2005}%
  \BibitemOpen
  \bibfield  {author} {\bibinfo {author} {\bibfnamefont {X.}~\bibnamefont
  {Li}}, \bibinfo {author} {\bibfnamefont {J.~C.}\ \bibnamefont {Tully}},
  \bibinfo {author} {\bibfnamefont {H.~B.}\ \bibnamefont {Schlegel}},\ and\
  \bibinfo {author} {\bibfnamefont {M.~J.}\ \bibnamefont {Frisch}},\ }\href
  {https://doi.org/10.1063/1.2008258} {\bibfield  {journal} {\bibinfo
  {journal} {J.~Chem.\ Phys.}\ }\textbf {\bibinfo {volume} {123}},\ \bibinfo
  {pages} {084106} (\bibinfo {year} {2005}{\natexlab{a}})}\BibitemShut
  {NoStop}%
\bibitem [{\citenamefont {Bastida}\ \emph {et~al.}(2008)\citenamefont
  {Bastida}, \citenamefont {Z{\'u}{\=n}iga}, \citenamefont {Requena},\ and\
  \citenamefont {Miguel}}]{Bastida_Miguel:2008}%
  \BibitemOpen
  \bibfield  {author} {\bibinfo {author} {\bibfnamefont {A.}~\bibnamefont
  {Bastida}}, \bibinfo {author} {\bibfnamefont {J.}~\bibnamefont
  {Z{\'u}{\=n}iga}}, \bibinfo {author} {\bibfnamefont {A.}~\bibnamefont
  {Requena}},\ and\ \bibinfo {author} {\bibfnamefont {B.}~\bibnamefont
  {Miguel}},\ }\href {https://doi.org/10.1063/1.2992617} {\bibfield  {journal}
  {\bibinfo  {journal} {J.~Chem.\ Phys.}\ }\textbf {\bibinfo {volume} {129}},\
  \bibinfo {pages} {154501} (\bibinfo {year} {2008})}\BibitemShut {NoStop}%
\bibitem [{\citenamefont {Vacher}\ \emph {et~al.}(2014)\citenamefont {Vacher},
  \citenamefont {Mendive-Tapia}, \citenamefont {Bearpark},\ and\ \citenamefont
  {Robb}}]{Vacher_Robb:2014}%
  \BibitemOpen
  \bibfield  {author} {\bibinfo {author} {\bibfnamefont {M.}~\bibnamefont
  {Vacher}}, \bibinfo {author} {\bibfnamefont {D.}~\bibnamefont
  {Mendive-Tapia}}, \bibinfo {author} {\bibfnamefont {M.~J.}\ \bibnamefont
  {Bearpark}},\ and\ \bibinfo {author} {\bibfnamefont {M.~A.}\ \bibnamefont
  {Robb}},\ }\href@noop {} {\bibfield  {journal} {\bibinfo  {journal} {Theor.\
  Chem.\ Acc.}\ }\textbf {\bibinfo {volume} {133}},\ \bibinfo {pages} {1}
  (\bibinfo {year} {2014})}\BibitemShut {NoStop}%
\bibitem [{\citenamefont {Donoso}\ and\ \citenamefont
  {Martens}(1998)}]{Donoso_Martens:1998}%
  \BibitemOpen
  \bibfield  {author} {\bibinfo {author} {\bibfnamefont {A.}~\bibnamefont
  {Donoso}}\ and\ \bibinfo {author} {\bibfnamefont {C.~C.}\ \bibnamefont
  {Martens}},\ }\href@noop {} {\bibfield  {journal} {\bibinfo  {journal}
  {J.~Phys.\ Chem.~A}\ }\textbf {\bibinfo {volume} {102}},\ \bibinfo {pages}
  {4291} (\bibinfo {year} {1998})}\BibitemShut {NoStop}%
\bibitem [{\citenamefont {Kapral}\ and\ \citenamefont
  {Ciccotti}(1999)}]{Kapral_Ciccotti:1999}%
  \BibitemOpen
  \bibfield  {author} {\bibinfo {author} {\bibfnamefont {R.}~\bibnamefont
  {Kapral}}\ and\ \bibinfo {author} {\bibfnamefont {G.}~\bibnamefont
  {Ciccotti}},\ }\href {https://doi.org/10.1063/1.478811} {\bibfield  {journal}
  {\bibinfo  {journal} {J.~Chem.\ Phys.}\ }\textbf {\bibinfo {volume} {110}},\
  \bibinfo {pages} {8919} (\bibinfo {year} {1999})}\BibitemShut {NoStop}%
\bibitem [{\citenamefont {Shi}\ and\ \citenamefont
  {Geva}(2004)}]{Shi_Geva:2004}%
  \BibitemOpen
  \bibfield  {author} {\bibinfo {author} {\bibfnamefont {Q.}~\bibnamefont
  {Shi}}\ and\ \bibinfo {author} {\bibfnamefont {E.}~\bibnamefont {Geva}},\
  }\href@noop {} {\bibfield  {journal} {\bibinfo  {journal} {J.~Chem.\ Phys.}\
  }\textbf {\bibinfo {volume} {121}},\ \bibinfo {pages} {3393} (\bibinfo {year}
  {2004})}\BibitemShut {NoStop}%
\bibitem [{\citenamefont {Meyer}\ and\ \citenamefont
  {Miller}(1979)}]{Meyer_Miller:1979}%
  \BibitemOpen
  \bibfield  {author} {\bibinfo {author} {\bibfnamefont {H.-D.}\ \bibnamefont
  {Meyer}}\ and\ \bibinfo {author} {\bibfnamefont {W.~H.}\ \bibnamefont
  {Miller}},\ }\href {https://doi.org/10.1063/1.437910} {\bibfield  {journal}
  {\bibinfo  {journal} {J.~Chem.\ Phys.}\ }\textbf {\bibinfo {volume} {70}},\
  \bibinfo {pages} {3214} (\bibinfo {year} {1979})}\BibitemShut {NoStop}%
\bibitem [{\citenamefont {Stock}\ and\ \citenamefont
  {Thoss}(1997)}]{Stock_Thoss:1997}%
  \BibitemOpen
  \bibfield  {author} {\bibinfo {author} {\bibfnamefont {G.}~\bibnamefont
  {Stock}}\ and\ \bibinfo {author} {\bibfnamefont {M.}~\bibnamefont {Thoss}},\
  }\href {https://doi.org/10.1103/PhysRevLett.78.578} {\bibfield  {journal}
  {\bibinfo  {journal} {Phys. Rev. Lett.}\ }\textbf {\bibinfo {volume} {78}},\
  \bibinfo {pages} {578} (\bibinfo {year} {1997})}\BibitemShut {NoStop}%
\bibitem [{\citenamefont {Miller}(2009)}]{Miller:2009}%
  \BibitemOpen
  \bibfield  {author} {\bibinfo {author} {\bibfnamefont {W.~H.}\ \bibnamefont
  {Miller}},\ }\href {https://doi.org/10.1021/jp809907p} {\bibfield  {journal}
  {\bibinfo  {journal} {J.~Phys.\ Chem.~A}\ }\textbf {\bibinfo {volume}
  {113}},\ \bibinfo {pages} {1405} (\bibinfo {year} {2009})}\BibitemShut
  {NoStop}%
\bibitem [{\citenamefont {Cotton}, \citenamefont {Liang},\ and\ \citenamefont
  {Miller}(2017)}]{Cotton_Miller:2017}%
  \BibitemOpen
  \bibfield  {author} {\bibinfo {author} {\bibfnamefont {S.~J.}\ \bibnamefont
  {Cotton}}, \bibinfo {author} {\bibfnamefont {R.}~\bibnamefont {Liang}},\ and\
  \bibinfo {author} {\bibfnamefont {W.~H.}\ \bibnamefont {Miller}},\ }\href
  {https://doi.org/10.1063/1.4995301} {\bibfield  {journal} {\bibinfo
  {journal} {J.~Chem.\ Phys.}\ }\textbf {\bibinfo {volume} {147}},\ \bibinfo
  {pages} {064112} (\bibinfo {year} {2017})}\BibitemShut {NoStop}%
\bibitem [{\citenamefont {Dunkel}, \citenamefont {Bonella},\ and\ \citenamefont
  {Coker}(2008)}]{Dunkel_Bonella:2008}%
  \BibitemOpen
  \bibfield  {author} {\bibinfo {author} {\bibfnamefont {E.~R.}\ \bibnamefont
  {Dunkel}}, \bibinfo {author} {\bibfnamefont {S.}~\bibnamefont {Bonella}},\
  and\ \bibinfo {author} {\bibfnamefont {D.~F.}\ \bibnamefont {Coker}},\ }\href
  {https://doi.org/10.1063/1.2976441} {\bibfield  {journal} {\bibinfo
  {journal} {J.~Chem.\ Phys.}\ }\textbf {\bibinfo {volume} {129}},\ \bibinfo
  {eid} {114106} (\bibinfo {year} {2008})}\BibitemShut {NoStop}%
\bibitem [{\citenamefont {Ananth}\ and\ \citenamefont
  {Miller}(2010)}]{Ananth_Miller:2010}%
  \BibitemOpen
  \bibfield  {author} {\bibinfo {author} {\bibfnamefont {N.}~\bibnamefont
  {Ananth}}\ and\ \bibinfo {author} {\bibfnamefont {T.~F.}\ \bibnamefont
  {Miller}},\ }\href {https://doi.org/10.1063/1.3511700} {\bibfield  {journal}
  {\bibinfo  {journal} {J.~Chem.\ Phys.}\ }\textbf {\bibinfo {volume} {133}},\
  \bibinfo {pages} {234103} (\bibinfo {year} {2010})}\BibitemShut {NoStop}%
\bibitem [{\citenamefont {Hele}\ and\ \citenamefont
  {Ananth}(2016)}]{Hele_Ananth:2016}%
  \BibitemOpen
  \bibfield  {author} {\bibinfo {author} {\bibfnamefont {T.~J.~H.}\
  \bibnamefont {Hele}}\ and\ \bibinfo {author} {\bibfnamefont {N.}~\bibnamefont
  {Ananth}},\ }\href {https://doi.org/10.1039/C6FD00106H} {\bibfield  {journal}
  {\bibinfo  {journal} {Faraday Discuss.}\ }\textbf {\bibinfo {volume} {195}},\
  \bibinfo {pages} {269} (\bibinfo {year} {2016})}\BibitemShut {NoStop}%
\bibitem [{\citenamefont {Domcke}\ and\ \citenamefont
  {Yarkony}(2012)}]{Domcke_Yarkony:2012}%
  \BibitemOpen
  \bibfield  {author} {\bibinfo {author} {\bibfnamefont {W.}~\bibnamefont
  {Domcke}}\ and\ \bibinfo {author} {\bibfnamefont {D.~R.}\ \bibnamefont
  {Yarkony}},\ }\href {https://doi.org/10.1146/annurev-physchem-032210-103522}
  {\bibfield  {journal} {\bibinfo  {journal} {Annu.\ Rev.\ Phys.\ Chem.}\
  }\textbf {\bibinfo {volume} {63}},\ \bibinfo {pages} {325} (\bibinfo {year}
  {2012})}\BibitemShut {NoStop}%
\bibitem [{\citenamefont {Takatsuka}\ \emph {et~al.}(2015)\citenamefont
  {Takatsuka}, \citenamefont {Yonehara}, \citenamefont {Hanasaki},\ and\
  \citenamefont {Arasaki}}]{book_Takatsuka:2015}%
  \BibitemOpen
  \bibfield  {author} {\bibinfo {author} {\bibfnamefont {K.}~\bibnamefont
  {Takatsuka}}, \bibinfo {author} {\bibfnamefont {T.}~\bibnamefont {Yonehara}},
  \bibinfo {author} {\bibfnamefont {K.}~\bibnamefont {Hanasaki}},\ and\
  \bibinfo {author} {\bibfnamefont {Y.}~\bibnamefont {Arasaki}},\ }\href@noop
  {} {\emph {\bibinfo {title} {Chemical Theory Beyond the Born-Oppenheimer
  Paradigm: Nonadiabatic Electronic and Nuclear Dynamics in Chemical
  Reactions}}}\ (\bibinfo  {publisher} {World Scientific},\ \bibinfo {address}
  {Singapore},\ \bibinfo {year} {2015})\BibitemShut {NoStop}%
\bibitem [{\citenamefont {Bircher}\ \emph {et~al.}(2017)\citenamefont
  {Bircher}, \citenamefont {Liberatore}, \citenamefont {Browning},
  \citenamefont {Brickel}, \citenamefont {Hofmann}, \citenamefont {Patoz},
  \citenamefont {Unke}, \citenamefont {Zimmermann}, \citenamefont {Chergui},
  \citenamefont {Hamm}, \citenamefont {Keller}, \citenamefont {Meuwly},
  \citenamefont {Woerner}, \citenamefont {Van{\'{i}}{\v{c}}ek},\ and\
  \citenamefont {Rothlisberger}}]{Bircher_Rothlisberger:2017}%
  \BibitemOpen
  \bibfield  {author} {\bibinfo {author} {\bibfnamefont {M.~P.}\ \bibnamefont
  {Bircher}}, \bibinfo {author} {\bibfnamefont {E.}~\bibnamefont {Liberatore}},
  \bibinfo {author} {\bibfnamefont {N.~J.}\ \bibnamefont {Browning}}, \bibinfo
  {author} {\bibfnamefont {S.}~\bibnamefont {Brickel}}, \bibinfo {author}
  {\bibfnamefont {C.}~\bibnamefont {Hofmann}}, \bibinfo {author} {\bibfnamefont
  {A.}~\bibnamefont {Patoz}}, \bibinfo {author} {\bibfnamefont {O.~T.}\
  \bibnamefont {Unke}}, \bibinfo {author} {\bibfnamefont {T.}~\bibnamefont
  {Zimmermann}}, \bibinfo {author} {\bibfnamefont {M.}~\bibnamefont {Chergui}},
  \bibinfo {author} {\bibfnamefont {P.}~\bibnamefont {Hamm}}, \bibinfo {author}
  {\bibfnamefont {U.}~\bibnamefont {Keller}}, \bibinfo {author} {\bibfnamefont
  {M.}~\bibnamefont {Meuwly}}, \bibinfo {author} {\bibfnamefont {H.~J.}\
  \bibnamefont {Woerner}}, \bibinfo {author} {\bibfnamefont {J.}~\bibnamefont
  {Van{\'{i}}{\v{c}}ek}},\ and\ \bibinfo {author} {\bibfnamefont
  {U.}~\bibnamefont {Rothlisberger}},\ }\href
  {https://doi.org/10.1063/1.4996816} {\bibfield  {journal} {\bibinfo
  {journal} {Struct. Dyn.}\ }\textbf {\bibinfo {volume} {4}},\ \bibinfo {pages}
  {061510} (\bibinfo {year} {2017})}\BibitemShut {NoStop}%
\bibitem [{\citenamefont {Zimmermann}\ and\ \citenamefont
  {Van\'{\i}\v{c}ek}(2010)}]{Zimmermann_Vanicek:2010}%
  \BibitemOpen
  \bibfield  {author} {\bibinfo {author} {\bibfnamefont {T.}~\bibnamefont
  {Zimmermann}}\ and\ \bibinfo {author} {\bibfnamefont {J.}~\bibnamefont
  {Van\'{\i}\v{c}ek}},\ }\href {https://doi.org/10.1063/1.3451266} {\bibfield
  {journal} {\bibinfo  {journal} {J.~Chem.\ Phys.}\ }\textbf {\bibinfo {volume}
  {132}},\ \bibinfo {eid} {241101} (\bibinfo {year} {2010})}\BibitemShut
  {NoStop}%
\bibitem [{\citenamefont {Zimmermann}\ and\ \citenamefont
  {Van\'{i}\v{c}ek}(2012{\natexlab{a}})}]{Zimmermann_Vanicek:2012}%
  \BibitemOpen
  \bibfield  {author} {\bibinfo {author} {\bibfnamefont {T.}~\bibnamefont
  {Zimmermann}}\ and\ \bibinfo {author} {\bibfnamefont {J.}~\bibnamefont
  {Van\'{i}\v{c}ek}},\ }\href {https://doi.org/10.1063/1.3690458} {\bibfield
  {journal} {\bibinfo  {journal} {J.~Chem.\ Phys.}\ }\textbf {\bibinfo {volume}
  {136}},\ \bibinfo {pages} {094106} (\bibinfo {year}
  {2012}{\natexlab{a}})}\BibitemShut {NoStop}%
\bibitem [{\citenamefont {Zimmermann}\ and\ \citenamefont
  {Van\'{i}\v{c}ek}(2012{\natexlab{b}})}]{Zimmermann_Vanicek:2012a}%
  \BibitemOpen
  \bibfield  {author} {\bibinfo {author} {\bibfnamefont {T.}~\bibnamefont
  {Zimmermann}}\ and\ \bibinfo {author} {\bibfnamefont {J.}~\bibnamefont
  {Van\'{i}\v{c}ek}},\ }\href {https://doi.org/10.1063/1.4738878} {\bibfield
  {journal} {\bibinfo  {journal} {J.~Chem.\ Phys.}\ }\textbf {\bibinfo {volume}
  {137}},\ \bibinfo {pages} {22A516} (\bibinfo {year}
  {2012}{\natexlab{b}})}\BibitemShut {NoStop}%
\bibitem [{\citenamefont {Parandekar}\ and\ \citenamefont
  {Tully}(2006)}]{Parandekar_Tully:2006}%
  \BibitemOpen
  \bibfield  {author} {\bibinfo {author} {\bibfnamefont {P.~V.}\ \bibnamefont
  {Parandekar}}\ and\ \bibinfo {author} {\bibfnamefont {J.~C.}\ \bibnamefont
  {Tully}},\ }\href {https://doi.org/10.1021/ct050213k} {\bibfield  {journal}
  {\bibinfo  {journal} {J.~Chem.\ Theory Comput.}\ }\textbf {\bibinfo {volume}
  {2}},\ \bibinfo {pages} {229} (\bibinfo {year} {2006})}\BibitemShut {NoStop}%
\bibitem [{\citenamefont {Loaiza}\ and\ \citenamefont
  {Izmaylov}(2018)}]{Loaiza_Izmaylov:2018}%
  \BibitemOpen
  \bibfield  {author} {\bibinfo {author} {\bibfnamefont {I.}~\bibnamefont
  {Loaiza}}\ and\ \bibinfo {author} {\bibfnamefont {A.~F.}\ \bibnamefont
  {Izmaylov}},\ }\href {https://doi.org/10.1063/1.5055768} {\bibfield
  {journal} {\bibinfo  {journal} {J.~Chem.\ Phys.}\ }\textbf {\bibinfo {volume}
  {149}},\ \bibinfo {pages} {214101} (\bibinfo {year} {2018})}\BibitemShut
  {NoStop}%
\bibitem [{\citenamefont {Blancafort}, \citenamefont {Hunt},\ and\
  \citenamefont {Robb}(2005)}]{Blancafort_Robb:2005}%
  \BibitemOpen
  \bibfield  {author} {\bibinfo {author} {\bibfnamefont {L.}~\bibnamefont
  {Blancafort}}, \bibinfo {author} {\bibfnamefont {P.}~\bibnamefont {Hunt}},\
  and\ \bibinfo {author} {\bibfnamefont {M.~A.}\ \bibnamefont {Robb}},\ }\href
  {https://doi.org/10.1021/ja043879h} {\bibfield  {journal} {\bibinfo
  {journal} {J.~Am.\ Chem.\ Soc.}\ }\textbf {\bibinfo {volume} {127}},\
  \bibinfo {pages} {3391} (\bibinfo {year} {2005})}\BibitemShut {NoStop}%
\bibitem [{\citenamefont {Wang}\ \emph {et~al.}(2011)\citenamefont {Wang},
  \citenamefont {Xu}, \citenamefont {Hong}, \citenamefont {Wang},\ and\
  \citenamefont {Gou}}]{Wang_Gou:2011}%
  \BibitemOpen
  \bibfield  {author} {\bibinfo {author} {\bibfnamefont {F.}~\bibnamefont
  {Wang}}, \bibinfo {author} {\bibfnamefont {X.}~\bibnamefont {Xu}}, \bibinfo
  {author} {\bibfnamefont {X.}~\bibnamefont {Hong}}, \bibinfo {author}
  {\bibfnamefont {J.}~\bibnamefont {Wang}},\ and\ \bibinfo {author}
  {\bibfnamefont {B.}~\bibnamefont {Gou}},\ }\href
  {https://doi.org/https://doi.org/10.1016/j.physleta.2011.07.032} {\bibfield
  {journal} {\bibinfo  {journal} {Phys.\ Lett.~A}\ }\textbf {\bibinfo {volume}
  {375}},\ \bibinfo {pages} {3290} (\bibinfo {year} {2011})}\BibitemShut
  {NoStop}%
\bibitem [{\citenamefont {Xie}\ \emph {et~al.}(2013)\citenamefont {Xie},
  \citenamefont {Bai}, \citenamefont {Zhu},\ and\ \citenamefont
  {Shi}}]{Xie_Qiang:2013}%
  \BibitemOpen
  \bibfield  {author} {\bibinfo {author} {\bibfnamefont {W.}~\bibnamefont
  {Xie}}, \bibinfo {author} {\bibfnamefont {S.}~\bibnamefont {Bai}}, \bibinfo
  {author} {\bibfnamefont {L.}~\bibnamefont {Zhu}},\ and\ \bibinfo {author}
  {\bibfnamefont {Q.}~\bibnamefont {Shi}},\ }\href
  {https://doi.org/10.1021/jp400462f} {\bibfield  {journal} {\bibinfo
  {journal} {J.~Phys.\ Chem.~A}\ }\textbf {\bibinfo {volume} {117}},\ \bibinfo
  {pages} {6196} (\bibinfo {year} {2013})}\BibitemShut {NoStop}%
\bibitem [{\citenamefont {Li}\ \emph {et~al.}(2013)\citenamefont {Li},
  \citenamefont {Movaghar}, \citenamefont {Nitzan},\ and\ \citenamefont
  {Ratner}}]{Li_Ratner:2013}%
  \BibitemOpen
  \bibfield  {author} {\bibinfo {author} {\bibfnamefont {G.}~\bibnamefont
  {Li}}, \bibinfo {author} {\bibfnamefont {B.}~\bibnamefont {Movaghar}},
  \bibinfo {author} {\bibfnamefont {A.}~\bibnamefont {Nitzan}},\ and\ \bibinfo
  {author} {\bibfnamefont {M.~A.}\ \bibnamefont {Ratner}},\ }\href
  {https://doi.org/10.1063/1.4776230} {\bibfield  {journal} {\bibinfo
  {journal} {J.~Chem.\ Phys.}\ }\textbf {\bibinfo {volume} {138}},\ \bibinfo
  {pages} {044112} (\bibinfo {year} {2013})}\BibitemShut {NoStop}%
\bibitem [{\citenamefont {Akimov}, \citenamefont {Long},\ and\ \citenamefont
  {Prezhdo}(2014)}]{Akimov_Prezhdo:2014}%
  \BibitemOpen
  \bibfield  {author} {\bibinfo {author} {\bibfnamefont {A.~V.}\ \bibnamefont
  {Akimov}}, \bibinfo {author} {\bibfnamefont {R.}~\bibnamefont {Long}},\ and\
  \bibinfo {author} {\bibfnamefont {O.~V.}\ \bibnamefont {Prezhdo}},\ }\href
  {https://doi.org/10.1063/1.4875702} {\bibfield  {journal} {\bibinfo
  {journal} {J.~Chem.\ Phys.}\ }\textbf {\bibinfo {volume} {140}},\ \bibinfo
  {pages} {194107} (\bibinfo {year} {2014})}\BibitemShut {NoStop}%
\bibitem [{\citenamefont {Kirson}\ \emph {et~al.}(1984)\citenamefont {Kirson},
  \citenamefont {Gerber}, \citenamefont {Nitzan},\ and\ \citenamefont
  {Ratner}}]{Kirson_Ratner:1984}%
  \BibitemOpen
  \bibfield  {author} {\bibinfo {author} {\bibfnamefont {Z.}~\bibnamefont
  {Kirson}}, \bibinfo {author} {\bibfnamefont {R.}~\bibnamefont {Gerber}},
  \bibinfo {author} {\bibfnamefont {A.}~\bibnamefont {Nitzan}},\ and\ \bibinfo
  {author} {\bibfnamefont {M.}~\bibnamefont {Ratner}},\ }\href
  {https://doi.org/https://doi.org/10.1016/0039-6028(84)90528-4} {\bibfield
  {journal} {\bibinfo  {journal} {Surface Science}\ }\textbf {\bibinfo {volume}
  {137}},\ \bibinfo {pages} {527} (\bibinfo {year} {1984})}\BibitemShut
  {NoStop}%
\bibitem [{\citenamefont {Kirson}\ \emph {et~al.}(1985)\citenamefont {Kirson},
  \citenamefont {Gerber}, \citenamefont {Nitzan},\ and\ \citenamefont
  {Ratner}}]{Kirson_Ratner:1985}%
  \BibitemOpen
  \bibfield  {author} {\bibinfo {author} {\bibfnamefont {Z.}~\bibnamefont
  {Kirson}}, \bibinfo {author} {\bibfnamefont {R.}~\bibnamefont {Gerber}},
  \bibinfo {author} {\bibfnamefont {A.}~\bibnamefont {Nitzan}},\ and\ \bibinfo
  {author} {\bibfnamefont {M.}~\bibnamefont {Ratner}},\ }\href
  {https://doi.org/https://doi.org/10.1016/0039-6028(85)90391-7} {\bibfield
  {journal} {\bibinfo  {journal} {Surface Science}\ }\textbf {\bibinfo {volume}
  {151}},\ \bibinfo {pages} {531} (\bibinfo {year} {1985})}\BibitemShut
  {NoStop}%
\bibitem [{\citenamefont {Head-Gordon}\ and\ \citenamefont
  {Tully}(1995)}]{Head-Gordon_Tully:1995}%
  \BibitemOpen
  \bibfield  {author} {\bibinfo {author} {\bibfnamefont {M.}~\bibnamefont
  {Head-Gordon}}\ and\ \bibinfo {author} {\bibfnamefont {J.~C.}\ \bibnamefont
  {Tully}},\ }\href {https://doi.org/10.1063/1.469915} {\bibfield  {journal}
  {\bibinfo  {journal} {J.~Chem.\ Phys.}\ }\textbf {\bibinfo {volume} {103}},\
  \bibinfo {pages} {10137} (\bibinfo {year} {1995})}\BibitemShut {NoStop}%
\bibitem [{\citenamefont {Ryabinkin}\ and\ \citenamefont
  {Izmaylov}(2017)}]{Ryabinkin_Izmaylov:2017}%
  \BibitemOpen
  \bibfield  {author} {\bibinfo {author} {\bibfnamefont {I.~G.}\ \bibnamefont
  {Ryabinkin}}\ and\ \bibinfo {author} {\bibfnamefont {A.~F.}\ \bibnamefont
  {Izmaylov}},\ }\href {https://doi.org/10.1021/acs.jpclett.6b02712} {\bibfield
   {journal} {\bibinfo  {journal} {J.~Phys.\ Chem.\ Lett.}\ }\textbf {\bibinfo
  {volume} {8}},\ \bibinfo {pages} {440} (\bibinfo {year} {2017})}\BibitemShut
  {NoStop}%
\bibitem [{\citenamefont {Topaler}\ \emph {et~al.}(1998)\citenamefont
  {Topaler}, \citenamefont {Allison}, \citenamefont {Schwenke},\ and\
  \citenamefont {Truhlar}}]{Topaler_Truhlar:1998}%
  \BibitemOpen
  \bibfield  {author} {\bibinfo {author} {\bibfnamefont {M.~S.}\ \bibnamefont
  {Topaler}}, \bibinfo {author} {\bibfnamefont {T.~C.}\ \bibnamefont
  {Allison}}, \bibinfo {author} {\bibfnamefont {D.~W.}\ \bibnamefont
  {Schwenke}},\ and\ \bibinfo {author} {\bibfnamefont {D.~G.}\ \bibnamefont
  {Truhlar}},\ }\href {https://doi.org/10.1063/1.477684} {\bibfield  {journal}
  {\bibinfo  {journal} {J.~Chem.\ Phys.}\ }\textbf {\bibinfo {volume} {109}},\
  \bibinfo {pages} {3321} (\bibinfo {year} {1998})}\BibitemShut {NoStop}%
\bibitem [{\citenamefont {Klein}\ \emph {et~al.}(1998)\citenamefont {Klein},
  \citenamefont {Bearpark}, \citenamefont {Smith}, \citenamefont {Robb},
  \citenamefont {Olivucci},\ and\ \citenamefont
  {Bernardi}}]{Klein_Bernardi:1998}%
  \BibitemOpen
  \bibfield  {author} {\bibinfo {author} {\bibfnamefont {S.}~\bibnamefont
  {Klein}}, \bibinfo {author} {\bibfnamefont {M.~J.}\ \bibnamefont {Bearpark}},
  \bibinfo {author} {\bibfnamefont {B.~R.}\ \bibnamefont {Smith}}, \bibinfo
  {author} {\bibfnamefont {M.~A.}\ \bibnamefont {Robb}}, \bibinfo {author}
  {\bibfnamefont {M.}~\bibnamefont {Olivucci}},\ and\ \bibinfo {author}
  {\bibfnamefont {F.}~\bibnamefont {Bernardi}},\ }\href
  {https://doi.org/10.1016/S0009-2614(98)00681-2} {\bibfield  {journal}
  {\bibinfo  {journal} {Chem.\ Phys.\ Lett.}\ }\textbf {\bibinfo {volume}
  {292}},\ \bibinfo {pages} {259} (\bibinfo {year} {1998})}\BibitemShut
  {NoStop}%
\bibitem [{\citenamefont {Gherib}, \citenamefont {Ryabinkin},\ and\
  \citenamefont {Izmaylov}(2015)}]{Gherib_Izmaylov:2015}%
  \BibitemOpen
  \bibfield  {author} {\bibinfo {author} {\bibfnamefont {R.}~\bibnamefont
  {Gherib}}, \bibinfo {author} {\bibfnamefont {I.~G.}\ \bibnamefont
  {Ryabinkin}},\ and\ \bibinfo {author} {\bibfnamefont {A.~F.}\ \bibnamefont
  {Izmaylov}},\ }\href {https://doi.org/10.1021/acs.jctc.5b00072} {\bibfield
  {journal} {\bibinfo  {journal} {J.~Chem.\ Theory Comput.}\ }\textbf {\bibinfo
  {volume} {11}},\ \bibinfo {pages} {1375} (\bibinfo {year}
  {2015})}\BibitemShut {NoStop}%
\bibitem [{\citenamefont {Kelly}, \citenamefont {Brackbill},\ and\
  \citenamefont {Markland}(2015)}]{Markland_Kelly:2015}%
  \BibitemOpen
  \bibfield  {author} {\bibinfo {author} {\bibfnamefont {A.}~\bibnamefont
  {Kelly}}, \bibinfo {author} {\bibfnamefont {N.}~\bibnamefont {Brackbill}},\
  and\ \bibinfo {author} {\bibfnamefont {T.~E.}\ \bibnamefont {Markland}},\
  }\href {https://doi.org/10.1063/1.4913686} {\bibfield  {journal} {\bibinfo
  {journal} {J.~Chem.\ Phys.}\ }\textbf {\bibinfo {volume} {142}},\ \bibinfo
  {pages} {094110} (\bibinfo {year} {2015})}\BibitemShut {NoStop}%
\bibitem [{\citenamefont {Zimmermann}\ and\ \citenamefont
  {Van{\'{i}}{\v{c}}ek}(2014)}]{Zimmermann_Vanicek:2014}%
  \BibitemOpen
  \bibfield  {author} {\bibinfo {author} {\bibfnamefont {T.}~\bibnamefont
  {Zimmermann}}\ and\ \bibinfo {author} {\bibfnamefont {J.}~\bibnamefont
  {Van{\'{i}}{\v{c}}ek}},\ }\href@noop {} {\bibfield  {journal} {\bibinfo
  {journal} {J.~Chem.\ Phys.}\ }\textbf {\bibinfo {volume} {141}},\ \bibinfo
  {pages} {134102} (\bibinfo {year} {2014})}\BibitemShut {NoStop}%
\bibitem [{\citenamefont {Ananth}, \citenamefont {Venkataraman},\ and\
  \citenamefont {Miller}(2007)}]{Ananth_Miller:2007}%
  \BibitemOpen
  \bibfield  {author} {\bibinfo {author} {\bibfnamefont {N.}~\bibnamefont
  {Ananth}}, \bibinfo {author} {\bibfnamefont {C.}~\bibnamefont
  {Venkataraman}},\ and\ \bibinfo {author} {\bibfnamefont {W.~H.}\ \bibnamefont
  {Miller}},\ }\href {https://doi.org/10.1063/1.2759932} {\bibfield  {journal}
  {\bibinfo  {journal} {J.~Chem.\ Phys.}\ }\textbf {\bibinfo {volume} {127}},\
  \bibinfo {pages} {084114} (\bibinfo {year} {2007})}\BibitemShut {NoStop}%
\bibitem [{\citenamefont {Shalashilin}(2009)}]{Shalashilin:2009}%
  \BibitemOpen
  \bibfield  {author} {\bibinfo {author} {\bibfnamefont {D.~V.}\ \bibnamefont
  {Shalashilin}},\ }\href {https://doi.org/10.1063/1.3153302} {\bibfield
  {journal} {\bibinfo  {journal} {J.~Chem.\ Phys.}\ }\textbf {\bibinfo {volume}
  {130}},\ \bibinfo {pages} {244101} (\bibinfo {year} {2009})}\BibitemShut
  {NoStop}%
\bibitem [{\citenamefont {Ma}\ \emph {et~al.}(2018)\citenamefont {Ma},
  \citenamefont {Bonfanti}, \citenamefont {Eisenbrandt}, \citenamefont
  {Martinazzo},\ and\ \citenamefont {Burghardt}}]{Ma_Burghardt:2018}%
  \BibitemOpen
  \bibfield  {author} {\bibinfo {author} {\bibfnamefont {T.}~\bibnamefont
  {Ma}}, \bibinfo {author} {\bibfnamefont {M.}~\bibnamefont {Bonfanti}},
  \bibinfo {author} {\bibfnamefont {P.}~\bibnamefont {Eisenbrandt}}, \bibinfo
  {author} {\bibfnamefont {R.}~\bibnamefont {Martinazzo}},\ and\ \bibinfo
  {author} {\bibfnamefont {I.}~\bibnamefont {Burghardt}},\ }\href
  {https://doi.org/10.1063/1.5062608} {\bibfield  {journal} {\bibinfo
  {journal} {J.~Chem.\ Phys.}\ }\textbf {\bibinfo {volume} {149}},\ \bibinfo
  {pages} {244107} (\bibinfo {year} {2018})}\BibitemShut {NoStop}%
\bibitem [{\citenamefont {Chen}\ \emph {et~al.}(2021)\citenamefont {Chen},
  \citenamefont {Sun}, \citenamefont {Shalashilin}, \citenamefont {Gelin},\
  and\ \citenamefont {Zhao}}]{Chen_Zhao:2021}%
  \BibitemOpen
  \bibfield  {author} {\bibinfo {author} {\bibfnamefont {L.}~\bibnamefont
  {Chen}}, \bibinfo {author} {\bibfnamefont {K.}~\bibnamefont {Sun}}, \bibinfo
  {author} {\bibfnamefont {D.~V.}\ \bibnamefont {Shalashilin}}, \bibinfo
  {author} {\bibfnamefont {M.~F.}\ \bibnamefont {Gelin}},\ and\ \bibinfo
  {author} {\bibfnamefont {Y.}~\bibnamefont {Zhao}},\ }\href
  {https://doi.org/10.1063/5.0038824} {\bibfield  {journal} {\bibinfo
  {journal} {J.~Chem.\ Phys.}\ }\textbf {\bibinfo {volume} {154}},\ \bibinfo
  {pages} {054105} (\bibinfo {year} {2021})}\BibitemShut {NoStop}%
\bibitem [{\citenamefont {Bornemann}, \citenamefont {Nettesheim},\ and\
  \citenamefont {Sch{\"u}tte}(1996)}]{Bornemann_Schutte:1996}%
  \BibitemOpen
  \bibfield  {author} {\bibinfo {author} {\bibfnamefont {F.~A.}\ \bibnamefont
  {Bornemann}}, \bibinfo {author} {\bibfnamefont {P.}~\bibnamefont
  {Nettesheim}},\ and\ \bibinfo {author} {\bibfnamefont {C.}~\bibnamefont
  {Sch{\"u}tte}},\ }\href {https://doi.org/10.1063/1.471952} {\bibfield
  {journal} {\bibinfo  {journal} {J.~Chem.\ Phys.}\ }\textbf {\bibinfo {volume}
  {105}},\ \bibinfo {pages} {1074} (\bibinfo {year} {1996})}\BibitemShut
  {NoStop}%
\bibitem [{\citenamefont {Feng}, \citenamefont {Micha},\ and\ \citenamefont
  {Runge}(1991)}]{Feng_Runge:1991}%
  \BibitemOpen
  \bibfield  {author} {\bibinfo {author} {\bibfnamefont {E.~Q.}\ \bibnamefont
  {Feng}}, \bibinfo {author} {\bibfnamefont {D.~A.}\ \bibnamefont {Micha}},\
  and\ \bibinfo {author} {\bibfnamefont {K.}~\bibnamefont {Runge}},\ }\href
  {https://doi.org/https://doi.org/10.1002/qua.560400409} {\bibfield  {journal}
  {\bibinfo  {journal} {Int.~J.~Quant.\ Chem.}\ }\textbf {\bibinfo {volume}
  {40}},\ \bibinfo {pages} {545} (\bibinfo {year} {1991})}\BibitemShut
  {NoStop}%
\bibitem [{\citenamefont {Ding}\ \emph {et~al.}(2015)\citenamefont {Ding},
  \citenamefont {Goings}, \citenamefont {Liu}, \citenamefont {Lingerfelt},\
  and\ \citenamefont {Li}}]{Ding_Li:2015}%
  \BibitemOpen
  \bibfield  {author} {\bibinfo {author} {\bibfnamefont {F.}~\bibnamefont
  {Ding}}, \bibinfo {author} {\bibfnamefont {J.~J.}\ \bibnamefont {Goings}},
  \bibinfo {author} {\bibfnamefont {H.}~\bibnamefont {Liu}}, \bibinfo {author}
  {\bibfnamefont {D.~B.}\ \bibnamefont {Lingerfelt}},\ and\ \bibinfo {author}
  {\bibfnamefont {X.}~\bibnamefont {Li}},\ }\href
  {https://doi.org/10.1063/1.4930985} {\bibfield  {journal} {\bibinfo
  {journal} {J.~Chem.\ Phys.}\ }\textbf {\bibinfo {volume} {143}},\ \bibinfo
  {pages} {114105} (\bibinfo {year} {2015})}\BibitemShut {NoStop}%
\bibitem [{\citenamefont {Hairer}, \citenamefont {Lubich},\ and\ \citenamefont
  {Wanner}(2006)}]{book_Hairer_Wanner:2006}%
  \BibitemOpen
  \bibfield  {author} {\bibinfo {author} {\bibfnamefont {E.}~\bibnamefont
  {Hairer}}, \bibinfo {author} {\bibfnamefont {C.}~\bibnamefont {Lubich}},\
  and\ \bibinfo {author} {\bibfnamefont {G.}~\bibnamefont {Wanner}},\ }\href
  {http://books.google.ch/books/about/Geometric_Numerical_Integration.html?id=T1TaNRLmZv8C&redir_esc=y}
  {\emph {\bibinfo {title} {Geometric Numerical Integration:
  Structure-Preserving Algorithms for Ordinary Differential Equations}}}\
  (\bibinfo  {publisher} {Springer Berlin Heidelberg New York},\ \bibinfo
  {year} {2006})\BibitemShut {NoStop}%
\bibitem [{\citenamefont {Leimkuhler}\ and\ \citenamefont
  {Reich}(2004)}]{book_Leimkuhler_Reich:2004}%
  \BibitemOpen
  \bibfield  {author} {\bibinfo {author} {\bibfnamefont {B.}~\bibnamefont
  {Leimkuhler}}\ and\ \bibinfo {author} {\bibfnamefont {S.}~\bibnamefont
  {Reich}},\ }\href
  {http://books.google.ch/books/about/Simulating_Hamiltonian_Dynamics.html?id=tpb-tnsZi5YC&redir_esc=y}
  {\emph {\bibinfo {title} {Simulating Hamiltonian Dynamics}}}\ (\bibinfo
  {publisher} {Cambridge University Press},\ \bibinfo {year}
  {2004})\BibitemShut {NoStop}%
\bibitem [{\citenamefont {Lubich}(2008)}]{book_Lubich:2008}%
  \BibitemOpen
  \bibfield  {author} {\bibinfo {author} {\bibfnamefont {C.}~\bibnamefont
  {Lubich}},\ }\href@noop {} {\emph {\bibinfo {title} {From Quantum to
  Classical Molecular Dynamics: Reduced Models and Numerical Analysis}}},\
  \bibinfo {edition} {12th}\ ed.\ (\bibinfo  {publisher} {European Mathematical
  Society},\ \bibinfo {address} {Z\"{u}rich},\ \bibinfo {year}
  {2008})\BibitemShut {NoStop}%
\bibitem [{\citenamefont {Nettesheim}\ \emph {et~al.}(1996)\citenamefont
  {Nettesheim}, \citenamefont {Bornemann}, \citenamefont {Schmidt},\ and\
  \citenamefont {Sch{\"u}tte}}]{Nettesheim_Schutte:1996}%
  \BibitemOpen
  \bibfield  {author} {\bibinfo {author} {\bibfnamefont {P.}~\bibnamefont
  {Nettesheim}}, \bibinfo {author} {\bibfnamefont {F.~A.}\ \bibnamefont
  {Bornemann}}, \bibinfo {author} {\bibfnamefont {B.}~\bibnamefont {Schmidt}},\
  and\ \bibinfo {author} {\bibfnamefont {C.}~\bibnamefont {Sch{\"u}tte}},\
  }\href {https://doi.org/https://doi.org/10.1016/0009-2614(96)00471-X}
  {\bibfield  {journal} {\bibinfo  {journal} {Chemical Physics Letters}\
  }\textbf {\bibinfo {volume} {256}},\ \bibinfo {pages} {581} (\bibinfo {year}
  {1996})}\BibitemShut {NoStop}%
\bibitem [{\citenamefont {Strang}(1968)}]{Strang:1968}%
  \BibitemOpen
  \bibfield  {author} {\bibinfo {author} {\bibfnamefont {G.}~\bibnamefont
  {Strang}},\ }\href {https://doi.org/10.1137/0705041} {\bibfield  {journal}
  {\bibinfo  {journal} {SIAM J.~Num.\ Analysis}\ }\textbf {\bibinfo {volume}
  {5}},\ \bibinfo {pages} {506} (\bibinfo {year} {1968})}\BibitemShut {NoStop}%
\bibitem [{\citenamefont {McLachlan}\ and\ \citenamefont
  {Quispel}(2002)}]{McLachlan_Quispel:2002}%
  \BibitemOpen
  \bibfield  {author} {\bibinfo {author} {\bibfnamefont {R.~I.}\ \bibnamefont
  {McLachlan}}\ and\ \bibinfo {author} {\bibfnamefont {G.~R.~W.}\ \bibnamefont
  {Quispel}},\ }\href@noop {} {\bibfield  {journal} {\bibinfo  {journal} {Acta
  Numerica}\ }\textbf {\bibinfo {volume} {11}},\ \bibinfo {pages} {341}
  (\bibinfo {year} {2002})}\BibitemShut {NoStop}%
\bibitem [{\citenamefont {Feit}, \citenamefont {Fleck},\ and\ \citenamefont
  {Steiger}(1982)}]{Feit_Steiger:1982}%
  \BibitemOpen
  \bibfield  {author} {\bibinfo {author} {\bibfnamefont {M.~D.}\ \bibnamefont
  {Feit}}, \bibinfo {author} {\bibfnamefont {J.~A.}\ \bibnamefont {Fleck},
  \bibfnamefont {{Jr.}}},\ and\ \bibinfo {author} {\bibfnamefont
  {A.}~\bibnamefont {Steiger}},\ }\href@noop {} {\bibfield  {journal} {\bibinfo
   {journal} {J.~Comp.\ Phys.}\ }\textbf {\bibinfo {volume} {47}},\ \bibinfo
  {pages} {412} (\bibinfo {year} {1982})}\BibitemShut {NoStop}%
\bibitem [{\citenamefont {Verlet}(1967)}]{Verlet:1967}%
  \BibitemOpen
  \bibfield  {author} {\bibinfo {author} {\bibfnamefont {L.}~\bibnamefont
  {Verlet}},\ }\href {https://doi.org/10.1103/PhysRev.159.98} {\bibfield
  {journal} {\bibinfo  {journal} {Phys. Rev.}\ }\textbf {\bibinfo {volume}
  {159}},\ \bibinfo {pages} {98} (\bibinfo {year} {1967})}\BibitemShut
  {NoStop}%
\bibitem [{\citenamefont {Fang}, \citenamefont {Jin},\ and\ \citenamefont
  {Sparber}(2018)}]{Fang_Sparber:2018}%
  \BibitemOpen
  \bibfield  {author} {\bibinfo {author} {\bibfnamefont {D.}~\bibnamefont
  {Fang}}, \bibinfo {author} {\bibfnamefont {S.}~\bibnamefont {Jin}},\ and\
  \bibinfo {author} {\bibfnamefont {C.}~\bibnamefont {Sparber}},\ }\href@noop
  {} {\bibfield  {journal} {\bibinfo  {journal} {Multiscale Model. Simul.}\
  }\textbf {\bibinfo {volume} {16}},\ \bibinfo {pages} {900} (\bibinfo {year}
  {2018})}\BibitemShut {NoStop}%
\bibitem [{\citenamefont {Kelly}\ \emph {et~al.}(2012)\citenamefont {Kelly},
  \citenamefont {van Zon}, \citenamefont {Schofield},\ and\ \citenamefont
  {Kapral}}]{Kelly_Kapral:2012}%
  \BibitemOpen
  \bibfield  {author} {\bibinfo {author} {\bibfnamefont {A.}~\bibnamefont
  {Kelly}}, \bibinfo {author} {\bibfnamefont {R.}~\bibnamefont {van Zon}},
  \bibinfo {author} {\bibfnamefont {J.}~\bibnamefont {Schofield}},\ and\
  \bibinfo {author} {\bibfnamefont {R.}~\bibnamefont {Kapral}},\ }\href
  {https://doi.org/10.1063/1.3685420} {\bibfield  {journal} {\bibinfo
  {journal} {J.~Chem.\ Phys.}\ }\textbf {\bibinfo {volume} {136}},\ \bibinfo
  {pages} {084101} (\bibinfo {year} {2012})}\BibitemShut {NoStop}%
\bibitem [{\citenamefont {Richardson}\ \emph {et~al.}(2017)\citenamefont
  {Richardson}, \citenamefont {Meyer}, \citenamefont {Pleinert},\ and\
  \citenamefont {Thoss}}]{Richardson_Thoss:2017}%
  \BibitemOpen
  \bibfield  {author} {\bibinfo {author} {\bibfnamefont {J.~O.}\ \bibnamefont
  {Richardson}}, \bibinfo {author} {\bibfnamefont {P.}~\bibnamefont {Meyer}},
  \bibinfo {author} {\bibfnamefont {M.-O.}\ \bibnamefont {Pleinert}},\ and\
  \bibinfo {author} {\bibfnamefont {M.}~\bibnamefont {Thoss}},\ }\href
  {https://doi.org/https://doi.org/10.1016/j.chemphys.2016.09.036} {\bibfield
  {journal} {\bibinfo  {journal} {Chem.\ Phys.}\ }\textbf {\bibinfo {volume}
  {482}},\ \bibinfo {pages} {124} (\bibinfo {year} {2017})}\BibitemShut
  {NoStop}%
\bibitem [{\citenamefont {Church}\ \emph {et~al.}(2018)\citenamefont {Church},
  \citenamefont {Hele}, \citenamefont {Ezra},\ and\ \citenamefont
  {Ananth}}]{Church_Ananth:2018}%
  \BibitemOpen
  \bibfield  {author} {\bibinfo {author} {\bibfnamefont {M.~S.}\ \bibnamefont
  {Church}}, \bibinfo {author} {\bibfnamefont {T.~J.~H.}\ \bibnamefont {Hele}},
  \bibinfo {author} {\bibfnamefont {G.~S.}\ \bibnamefont {Ezra}},\ and\
  \bibinfo {author} {\bibfnamefont {N.}~\bibnamefont {Ananth}},\ }\href
  {https://doi.org/10.1063/1.5005557} {\bibfield  {journal} {\bibinfo
  {journal} {J.~Chem.\ Phys.}\ }\textbf {\bibinfo {volume} {148}},\ \bibinfo
  {pages} {102326} (\bibinfo {year} {2018})}\BibitemShut {NoStop}%
\bibitem [{\citenamefont {Runeson}\ and\ \citenamefont
  {Richardson}(2020)}]{Runeson_Richardson:2020}%
  \BibitemOpen
  \bibfield  {author} {\bibinfo {author} {\bibfnamefont {J.~E.}\ \bibnamefont
  {Runeson}}\ and\ \bibinfo {author} {\bibfnamefont {J.~O.}\ \bibnamefont
  {Richardson}},\ }\href {https://doi.org/10.1063/1.5143412} {\bibfield
  {journal} {\bibinfo  {journal} {J.~Chem.\ Phys.}\ }\textbf {\bibinfo {volume}
  {152}},\ \bibinfo {pages} {084110} (\bibinfo {year} {2020})}\BibitemShut
  {NoStop}%
\bibitem [{\citenamefont {Goings}, \citenamefont {Lestrange},\ and\
  \citenamefont {Li}(2018)}]{Goings_Li:2018}%
  \BibitemOpen
  \bibfield  {author} {\bibinfo {author} {\bibfnamefont {J.~J.}\ \bibnamefont
  {Goings}}, \bibinfo {author} {\bibfnamefont {P.~J.}\ \bibnamefont
  {Lestrange}},\ and\ \bibinfo {author} {\bibfnamefont {X.}~\bibnamefont
  {Li}},\ }\href {https://doi.org/10.1002/wcms.1341} {\bibfield  {journal}
  {\bibinfo  {journal} {Wiley~Interdiscip.~Rev.~Comput.~Mol.~Sci.}\ }\textbf
  {\bibinfo {volume} {8}},\ \bibinfo {pages} {e1341} (\bibinfo {year}
  {2018})}\BibitemShut {NoStop}%
\bibitem [{\citenamefont {Li}\ \emph {et~al.}(2020)\citenamefont {Li},
  \citenamefont {Govind}, \citenamefont {Isborn}, \citenamefont {DePrince},\
  and\ \citenamefont {Lopata}}]{Li_Lopata:2020}%
  \BibitemOpen
  \bibfield  {author} {\bibinfo {author} {\bibfnamefont {X.}~\bibnamefont
  {Li}}, \bibinfo {author} {\bibfnamefont {N.}~\bibnamefont {Govind}}, \bibinfo
  {author} {\bibfnamefont {C.}~\bibnamefont {Isborn}}, \bibinfo {author}
  {\bibfnamefont {A.~E.}\ \bibnamefont {DePrince}},\ and\ \bibinfo {author}
  {\bibfnamefont {K.}~\bibnamefont {Lopata}},\ }\href
  {https://doi.org/10.1021/acs.chemrev.0c00223} {\bibfield  {journal} {\bibinfo
   {journal} {Chem.\ Rev.}\ }\textbf {\bibinfo {volume} {120}},\ \bibinfo
  {pages} {9951} (\bibinfo {year} {2020})}\BibitemShut {NoStop}%
\bibitem [{\citenamefont {Yoshida}(1990)}]{Yoshida:1990}%
  \BibitemOpen
  \bibfield  {author} {\bibinfo {author} {\bibfnamefont {H.}~\bibnamefont
  {Yoshida}},\ }\href {https://doi.org/10.1016/0375-9601(90)90092-3} {\bibfield
   {journal} {\bibinfo  {journal} {Phys.\ Lett.~A}\ }\textbf {\bibinfo {volume}
  {150}},\ \bibinfo {pages} {262} (\bibinfo {year} {1990})}\BibitemShut
  {NoStop}%
\bibitem [{\citenamefont {Suzuki}(1990)}]{Suzuki:1990}%
  \BibitemOpen
  \bibfield  {author} {\bibinfo {author} {\bibfnamefont {M.}~\bibnamefont
  {Suzuki}},\ }\href {https://doi.org/10.1016/0375-9601(90)90962-n} {\bibfield
  {journal} {\bibinfo  {journal} {Phys.\ Lett.~A}\ }\textbf {\bibinfo {volume}
  {146}},\ \bibinfo {pages} {319} (\bibinfo {year} {1990})}\BibitemShut
  {NoStop}%
\bibitem [{\citenamefont {McLachlan}(1995)}]{McLachlan:1995}%
  \BibitemOpen
  \bibfield  {author} {\bibinfo {author} {\bibfnamefont {R.~I.}\ \bibnamefont
  {McLachlan}},\ }\href {https://doi.org/10.1137/0916010} {\bibfield  {journal}
  {\bibinfo  {journal} {SIAM J.\ Sci.\ Comp.}\ }\textbf {\bibinfo {volume}
  {16}},\ \bibinfo {pages} {151} (\bibinfo {year} {1995})}\BibitemShut
  {NoStop}%
\bibitem [{\citenamefont {Kahan}\ and\ \citenamefont
  {Li}(1997)}]{Kahan_Li:1997}%
  \BibitemOpen
  \bibfield  {author} {\bibinfo {author} {\bibfnamefont {W.}~\bibnamefont
  {Kahan}}\ and\ \bibinfo {author} {\bibfnamefont {R.-C.}\ \bibnamefont {Li}},\
  }\href {https://doi.org/10.1090/s0025-5718-97-00873-9} {\bibfield  {journal}
  {\bibinfo  {journal} {Math.\ Comput.}\ }\textbf {\bibinfo {volume} {66}},\
  \bibinfo {pages} {1089} (\bibinfo {year} {1997})}\BibitemShut {NoStop}%
\bibitem [{\citenamefont {Sofroniou}\ and\ \citenamefont
  {Spaletta}(2005)}]{Sofroniou_Spaletta:2005}%
  \BibitemOpen
  \bibfield  {author} {\bibinfo {author} {\bibfnamefont {M.}~\bibnamefont
  {Sofroniou}}\ and\ \bibinfo {author} {\bibfnamefont {G.}~\bibnamefont
  {Spaletta}},\ }\href {https://doi.org/10.1080/10556780500140664} {\bibfield
  {journal} {\bibinfo  {journal} {Optim.\ Method Softw.}\ }\textbf {\bibinfo
  {volume} {20}},\ \bibinfo {pages} {597} (\bibinfo {year} {2005})}\BibitemShut
  {NoStop}%
\bibitem [{\citenamefont {Hader}\ \emph {et~al.}(2017)\citenamefont {Hader},
  \citenamefont {Albert}, \citenamefont {Gross},\ and\ \citenamefont
  {Engel}}]{Hader_Engel:2017}%
  \BibitemOpen
  \bibfield  {author} {\bibinfo {author} {\bibfnamefont {K.}~\bibnamefont
  {Hader}}, \bibinfo {author} {\bibfnamefont {J.}~\bibnamefont {Albert}},
  \bibinfo {author} {\bibfnamefont {E.~K.~U.}\ \bibnamefont {Gross}},\ and\
  \bibinfo {author} {\bibfnamefont {V.}~\bibnamefont {Engel}},\ }\href
  {https://doi.org/10.1063/1.4975811} {\bibfield  {journal} {\bibinfo
  {journal} {J.~Chem.\ Phys.}\ }\textbf {\bibinfo {volume} {146}},\ \bibinfo
  {pages} {074304} (\bibinfo {year} {2017})}\BibitemShut {NoStop}%
\bibitem [{\citenamefont {Albert}, \citenamefont {Hader},\ and\ \citenamefont
  {Engel}(2017)}]{Albert_Engel:2017}%
  \BibitemOpen
  \bibfield  {author} {\bibinfo {author} {\bibfnamefont {J.}~\bibnamefont
  {Albert}}, \bibinfo {author} {\bibfnamefont {K.}~\bibnamefont {Hader}},\ and\
  \bibinfo {author} {\bibfnamefont {V.}~\bibnamefont {Engel}},\ }\href
  {https://doi.org/10.1063/1.4989780} {\bibfield  {journal} {\bibinfo
  {journal} {J.~Chem.\ Phys.}\ }\textbf {\bibinfo {volume} {147}},\ \bibinfo
  {pages} {064302} (\bibinfo {year} {2017})}\BibitemShut {NoStop}%
\bibitem [{\citenamefont {Schaupp}\ and\ \citenamefont
  {Engel}(2019)}]{Schaupp_Engel:2019}%
  \BibitemOpen
  \bibfield  {author} {\bibinfo {author} {\bibfnamefont {T.}~\bibnamefont
  {Schaupp}}\ and\ \bibinfo {author} {\bibfnamefont {V.}~\bibnamefont
  {Engel}},\ }\href {https://doi.org/10.1063/1.5111922} {\bibfield  {journal}
  {\bibinfo  {journal} {J.~Chem.\ Phys.}\ }\textbf {\bibinfo {volume} {151}},\
  \bibinfo {pages} {084309} (\bibinfo {year} {2019})}\BibitemShut {NoStop}%
\bibitem [{\citenamefont {Shin}\ and\ \citenamefont
  {Metiu}(1995)}]{Shin_Metiu:1995}%
  \BibitemOpen
  \bibfield  {author} {\bibinfo {author} {\bibfnamefont {S.}~\bibnamefont
  {Shin}}\ and\ \bibinfo {author} {\bibfnamefont {H.}~\bibnamefont {Metiu}},\
  }\href@noop {} {\bibfield  {journal} {\bibinfo  {journal} {J.~Chem.\ Phys.}\
  }\textbf {\bibinfo {volume} {102}},\ \bibinfo {pages} {9285} (\bibinfo {year}
  {1995})}\BibitemShut {NoStop}%
\bibitem [{\citenamefont {Shin}\ and\ \citenamefont
  {Metiu}(1996)}]{Shin_Metiu:1996}%
  \BibitemOpen
  \bibfield  {author} {\bibinfo {author} {\bibfnamefont {S.}~\bibnamefont
  {Shin}}\ and\ \bibinfo {author} {\bibfnamefont {H.}~\bibnamefont {Metiu}},\
  }\href {https://doi.org/10.1021/jp952498a} {\bibfield  {journal} {\bibinfo
  {journal} {J.~Phys.~C}\ }\textbf {\bibinfo {volume} {100}},\ \bibinfo {pages}
  {7867} (\bibinfo {year} {1996})}\BibitemShut {NoStop}%
\bibitem [{\citenamefont {Heller}(1976)}]{Heller:1976}%
  \BibitemOpen
  \bibfield  {author} {\bibinfo {author} {\bibfnamefont {E.~J.}\ \bibnamefont
  {Heller}},\ }\href {https://doi.org/10.1063/1.431911} {\bibfield  {journal}
  {\bibinfo  {journal} {J.~Chem.\ Phys.}\ }\textbf {\bibinfo {volume} {64}},\
  \bibinfo {pages} {63} (\bibinfo {year} {1976})}\BibitemShut {NoStop}%
\bibitem [{\citenamefont {Gerber}, \citenamefont {Buch},\ and\ \citenamefont
  {Ratner}(1982)}]{Gerber_Ratner:1982}%
  \BibitemOpen
  \bibfield  {author} {\bibinfo {author} {\bibfnamefont {R.~B.}\ \bibnamefont
  {Gerber}}, \bibinfo {author} {\bibfnamefont {V.}~\bibnamefont {Buch}},\ and\
  \bibinfo {author} {\bibfnamefont {M.~A.}\ \bibnamefont {Ratner}},\ }\href
  {https://doi.org/10.1063/1.444225} {\bibfield  {journal} {\bibinfo  {journal}
  {J.~Chem.\ Phys.}\ }\textbf {\bibinfo {volume} {77}},\ \bibinfo {pages}
  {3022} (\bibinfo {year} {1982})}\BibitemShut {NoStop}%
\bibitem [{\citenamefont {Gerber}, \citenamefont {Ratner},\ and\ \citenamefont
  {Buch}(1982)}]{Gerber_Buch:1982}%
  \BibitemOpen
  \bibfield  {author} {\bibinfo {author} {\bibfnamefont {R.}~\bibnamefont
  {Gerber}}, \bibinfo {author} {\bibfnamefont {M.}~\bibnamefont {Ratner}},\
  and\ \bibinfo {author} {\bibfnamefont {V.}~\bibnamefont {Buch}},\ }\href
  {https://doi.org/https://doi.org/10.1016/0009-2614(82)83635-X} {\bibfield
  {journal} {\bibinfo  {journal} {Chem.\ Phys.\ Lett.}\ }\textbf {\bibinfo
  {volume} {91}},\ \bibinfo {pages} {173} (\bibinfo {year} {1982})}\BibitemShut
  {NoStop}%
\bibitem [{\citenamefont {Bisseling}\ \emph {et~al.}(1987)\citenamefont
  {Bisseling}, \citenamefont {Kosloff}, \citenamefont {Gerber}, \citenamefont
  {Ratner}, \citenamefont {Gibson},\ and\ \citenamefont
  {Cerjan}}]{Bisseling_Cerjan:1987}%
  \BibitemOpen
  \bibfield  {author} {\bibinfo {author} {\bibfnamefont {R.~H.}\ \bibnamefont
  {Bisseling}}, \bibinfo {author} {\bibfnamefont {R.}~\bibnamefont {Kosloff}},
  \bibinfo {author} {\bibfnamefont {R.~B.}\ \bibnamefont {Gerber}}, \bibinfo
  {author} {\bibfnamefont {M.~A.}\ \bibnamefont {Ratner}}, \bibinfo {author}
  {\bibfnamefont {L.}~\bibnamefont {Gibson}},\ and\ \bibinfo {author}
  {\bibfnamefont {C.}~\bibnamefont {Cerjan}},\ }\href
  {https://doi.org/10.1063/1.453063} {\bibfield  {journal} {\bibinfo  {journal}
  {J.~Chem.\ Phys.}\ }\textbf {\bibinfo {volume} {87}},\ \bibinfo {pages}
  {2760} (\bibinfo {year} {1987})}\BibitemShut {NoStop}%
\bibitem [{\citenamefont {Messina}\ and\ \citenamefont
  {Coalson}(1989)}]{Messina_Coalson:1989}%
  \BibitemOpen
  \bibfield  {author} {\bibinfo {author} {\bibfnamefont {M.}~\bibnamefont
  {Messina}}\ and\ \bibinfo {author} {\bibfnamefont {R.~D.}\ \bibnamefont
  {Coalson}},\ }\href {https://doi.org/10.1063/1.455812} {\bibfield  {journal}
  {\bibinfo  {journal} {J.~Chem.\ Phys.}\ }\textbf {\bibinfo {volume} {90}},\
  \bibinfo {pages} {4015} (\bibinfo {year} {1989})}\BibitemShut {NoStop}%
\bibitem [{\citenamefont {Dirac}(1930)}]{Dirac:1930}%
  \BibitemOpen
  \bibfield  {author} {\bibinfo {author} {\bibfnamefont {P.~A.~M.}\
  \bibnamefont {Dirac}},\ }\href {https://doi.org/10.1017/S0305004100016108}
  {\bibfield  {journal} {\bibinfo  {journal} {Math.\ Proc.\ Camb.\ Phil.\
  Soc.}\ }\textbf {\bibinfo {volume} {26}},\ \bibinfo {pages} {376–385}
  (\bibinfo {year} {1930})}\BibitemShut {NoStop}%
\bibitem [{\citenamefont {Frenkel}(1934)}]{book_Frenkel:1934}%
  \BibitemOpen
  \bibfield  {author} {\bibinfo {author} {\bibfnamefont {J.}~\bibnamefont
  {Frenkel}},\ }\href@noop {} {\emph {\bibinfo {title} {Wave mechanics}}}\
  (\bibinfo  {publisher} {Clarendon Press},\ \bibinfo {address} {Oxford},\
  \bibinfo {year} {1934})\BibitemShut {NoStop}%
\bibitem [{\citenamefont {Marsden}\ and\ \citenamefont
  {Ratiu}(1999)}]{book_Marsden_Ratiu:1999}%
  \BibitemOpen
  \bibfield  {author} {\bibinfo {author} {\bibfnamefont {J.~E.}\ \bibnamefont
  {Marsden}}\ and\ \bibinfo {author} {\bibfnamefont {T.~S.}\ \bibnamefont
  {Ratiu}},\ }\href@noop {} {\emph {\bibinfo {title} {Introduction to mechanics
  and symmetry: a basic exposition of classical mechanical systems}}},\
  Vol.~\bibinfo {volume} {17}\ (\bibinfo  {publisher} {Springer Science \&
  Business Media},\ \bibinfo {year} {1999})\BibitemShut {NoStop}%
\bibitem [{\citenamefont {Ohsawa}\ and\ \citenamefont
  {Leok}(2013)}]{Ohsawa_Leok:2013}%
  \BibitemOpen
  \bibfield  {author} {\bibinfo {author} {\bibfnamefont {T.}~\bibnamefont
  {Ohsawa}}\ and\ \bibinfo {author} {\bibfnamefont {M.}~\bibnamefont {Leok}},\
  }\href {https://doi.org/10.1088/1751-8113/46/40/405201} {\bibfield  {journal}
  {\bibinfo  {journal} {J.~Phys.~A}\ }\textbf {\bibinfo {volume} {46}},\
  \bibinfo {pages} {405201} (\bibinfo {year} {2013})}\BibitemShut {NoStop}%
\bibitem [{\citenamefont {Abraham}\ and\ \citenamefont
  {Marsden}(1978)}]{book_Abraham_Marsden:1978}%
  \BibitemOpen
  \bibfield  {author} {\bibinfo {author} {\bibfnamefont {R.}~\bibnamefont
  {Abraham}}\ and\ \bibinfo {author} {\bibfnamefont {J.~E.}\ \bibnamefont
  {Marsden}},\ }\href@noop {} {\emph {\bibinfo {title} {Foundations of
  mechanics}}}\ (\bibinfo  {publisher} {Addison-Wesley Publishing Company,
  Inc.},\ \bibinfo {year} {1978})\BibitemShut {NoStop}%
\bibitem [{\citenamefont {Wehrle}, \citenamefont {\v{S}ulc},\ and\
  \citenamefont {Van\'{\i}\v{c}ek}(2011)}]{Wehrle_Vanicek:2011}%
  \BibitemOpen
  \bibfield  {author} {\bibinfo {author} {\bibfnamefont {M.}~\bibnamefont
  {Wehrle}}, \bibinfo {author} {\bibfnamefont {M.}~\bibnamefont {\v{S}ulc}},\
  and\ \bibinfo {author} {\bibfnamefont {J.}~\bibnamefont {Van\'{\i}\v{c}ek}},\
  }\href {https://doi.org/10.2533/chimia.2011.334} {\bibfield  {journal}
  {\bibinfo  {journal} {Chimia}\ }\textbf {\bibinfo {volume} {65}},\ \bibinfo
  {pages} {334} (\bibinfo {year} {2011})}\BibitemShut {NoStop}%
\bibitem [{\citenamefont {Roulet}, \citenamefont {Choi},\ and\ \citenamefont
  {Van\'{i}\v{c}ek}(2019)}]{Roulet_Vanicek:2019}%
  \BibitemOpen
  \bibfield  {author} {\bibinfo {author} {\bibfnamefont {J.}~\bibnamefont
  {Roulet}}, \bibinfo {author} {\bibfnamefont {S.}~\bibnamefont {Choi}},\ and\
  \bibinfo {author} {\bibfnamefont {J.}~\bibnamefont {Van\'{i}\v{c}ek}},\
  }\href {https://doi.org/10.1063/1.5094046} {\bibfield  {journal} {\bibinfo
  {journal} {J.~Chem.\ Phys.}\ }\textbf {\bibinfo {volume} {150}},\ \bibinfo
  {pages} {204113} (\bibinfo {year} {2019})}\BibitemShut {NoStop}%
\bibitem [{\citenamefont {Choi}\ and\ \citenamefont
  {Van{\'{i}}{\v{c}}ek}(2019)}]{Choi_Vanicek:2019a}%
  \BibitemOpen
  \bibfield  {author} {\bibinfo {author} {\bibfnamefont {S.}~\bibnamefont
  {Choi}}\ and\ \bibinfo {author} {\bibfnamefont {J.}~\bibnamefont
  {Van{\'{i}}{\v{c}}ek}},\ }\href {https://doi.org/10.1063/1.5127856}
  {\bibfield  {journal} {\bibinfo  {journal} {J.~Chem.\ Phys.}\ }\textbf
  {\bibinfo {volume} {151}},\ \bibinfo {pages} {234102} (\bibinfo {year}
  {2019})}\BibitemShut {NoStop}%
\bibitem [{\citenamefont {Choi}\ and\ \citenamefont
  {Van\'{i}\v{c}ek}(2019)}]{Choi_Vanicek:2019}%
  \BibitemOpen
  \bibfield  {author} {\bibinfo {author} {\bibfnamefont {S.}~\bibnamefont
  {Choi}}\ and\ \bibinfo {author} {\bibfnamefont {J.}~\bibnamefont
  {Van\'{i}\v{c}ek}},\ }\href {https://doi.org/10.1063/1.5092611} {\bibfield
  {journal} {\bibinfo  {journal} {J.~Chem.\ Phys.}\ }\textbf {\bibinfo {volume}
  {150}},\ \bibinfo {pages} {204112} (\bibinfo {year} {2019})}\BibitemShut
  {NoStop}%
\bibitem [{\citenamefont {Nakamura}(2012)}]{book_Nakamura:2012}%
  \BibitemOpen
  \bibfield  {author} {\bibinfo {author} {\bibfnamefont {H.}~\bibnamefont
  {Nakamura}},\ }\href@noop {} {\emph {\bibinfo {title} {Nonadiabatic
  Transition: Concepts, Basic Theories and Applications}}},\ \bibinfo {edition}
  {2nd}\ ed.\ (\bibinfo  {publisher} {World Scientific Publishing Company},\
  \bibinfo {year} {2012})\BibitemShut {NoStop}%
\bibitem [{\citenamefont {Mukamel}(1999)}]{book_Mukamel:1999}%
  \BibitemOpen
  \bibfield  {author} {\bibinfo {author} {\bibfnamefont {S.}~\bibnamefont
  {Mukamel}},\ }\href@noop {} {\emph {\bibinfo {title} {Principles of nonlinear
  optical spectroscopy}}},\ \bibinfo {edition} {1st}\ ed.\ (\bibinfo
  {publisher} {Oxford University Press},\ \bibinfo {address} {New York},\
  \bibinfo {year} {1999})\BibitemShut {NoStop}%
\bibitem [{\citenamefont {Domcke}, \citenamefont {Yarkony},\ and\ \citenamefont
  {K{\"o}ppel}(2004)}]{book_Domcke_Koppel:2004}%
  \BibitemOpen
  \bibfield  {author} {\bibinfo {author} {\bibfnamefont {W.}~\bibnamefont
  {Domcke}}, \bibinfo {author} {\bibfnamefont {D.}~\bibnamefont {Yarkony}},\
  and\ \bibinfo {author} {\bibfnamefont {H.}~\bibnamefont {K{\"o}ppel}},\
  }\href@noop {} {\emph {\bibinfo {title} {Conical intersections: electronic
  structure, dynamics \& spectroscopy}}},\ Vol.~\bibinfo {volume} {15}\
  (\bibinfo  {publisher} {World Scientific},\ \bibinfo {year}
  {2004})\BibitemShut {NoStop}%
\bibitem [{\citenamefont {Heller}(2018)}]{book_Heller:2018}%
  \BibitemOpen
  \bibfield  {author} {\bibinfo {author} {\bibfnamefont {E.~J.}\ \bibnamefont
  {Heller}},\ }\href@noop {} {\emph {\bibinfo {title} {The semiclassical way to
  dynamics and spectroscopy}}}\ (\bibinfo  {publisher} {Princeton University
  Press},\ \bibinfo {address} {Princeton, NJ},\ \bibinfo {year}
  {2018})\BibitemShut {NoStop}%
\bibitem [{\citenamefont {Li}\ \emph {et~al.}(2005{\natexlab{b}})\citenamefont
  {Li}, \citenamefont {Smith}, \citenamefont {Markevitch}, \citenamefont
  {Romanov}, \citenamefont {Levis},\ and\ \citenamefont
  {Schlegel}}]{Li_Schlegel:2005}%
  \BibitemOpen
  \bibfield  {author} {\bibinfo {author} {\bibfnamefont {X.}~\bibnamefont
  {Li}}, \bibinfo {author} {\bibfnamefont {S.~M.}\ \bibnamefont {Smith}},
  \bibinfo {author} {\bibfnamefont {A.~N.}\ \bibnamefont {Markevitch}},
  \bibinfo {author} {\bibfnamefont {D.~A.}\ \bibnamefont {Romanov}}, \bibinfo
  {author} {\bibfnamefont {R.~J.}\ \bibnamefont {Levis}},\ and\ \bibinfo
  {author} {\bibfnamefont {H.~B.}\ \bibnamefont {Schlegel}},\ }\href
  {https://doi.org/10.1039/B415849K} {\bibfield  {journal} {\bibinfo  {journal}
  {Phys.\ Chem.\ Chem.\ Phys.}\ }\textbf {\bibinfo {volume} {7}},\ \bibinfo
  {pages} {233} (\bibinfo {year} {2005}{\natexlab{b}})}\BibitemShut {NoStop}%
\bibitem [{\citenamefont {Theilhaber}(1992)}]{Theilhaber:1992}%
  \BibitemOpen
  \bibfield  {author} {\bibinfo {author} {\bibfnamefont {J.}~\bibnamefont
  {Theilhaber}},\ }\href {https://doi.org/10.1103/PhysRevB.46.12990} {\bibfield
   {journal} {\bibinfo  {journal} {Phys.\ Rev.~B}\ }\textbf {\bibinfo {volume}
  {46}},\ \bibinfo {pages} {12990} (\bibinfo {year} {1992})}\BibitemShut
  {NoStop}%
\bibitem [{\citenamefont {Yabana}\ and\ \citenamefont
  {Bertsch}(1996)}]{Yabana_Bertsch:1996}%
  \BibitemOpen
  \bibfield  {author} {\bibinfo {author} {\bibfnamefont {K.}~\bibnamefont
  {Yabana}}\ and\ \bibinfo {author} {\bibfnamefont {G.~F.}\ \bibnamefont
  {Bertsch}},\ }\href {https://doi.org/10.1103/PhysRevB.54.4484} {\bibfield
  {journal} {\bibinfo  {journal} {Phys.\ Rev.~B}\ }\textbf {\bibinfo {volume}
  {54}},\ \bibinfo {pages} {4484} (\bibinfo {year} {1996})}\BibitemShut
  {NoStop}%
\bibitem [{\citenamefont {Castro}\ \emph {et~al.}(2004)\citenamefont {Castro},
  \citenamefont {Marques}, \citenamefont {Alonso}, \citenamefont {Bertsch},\
  and\ \citenamefont {Rubio}}]{Castro_Rubio:2004}%
  \BibitemOpen
  \bibfield  {author} {\bibinfo {author} {\bibfnamefont {A.}~\bibnamefont
  {Castro}}, \bibinfo {author} {\bibfnamefont {M.~A.}\ \bibnamefont {Marques}},
  \bibinfo {author} {\bibfnamefont {J.~A.}\ \bibnamefont {Alonso}}, \bibinfo
  {author} {\bibfnamefont {G.~F.}\ \bibnamefont {Bertsch}},\ and\ \bibinfo
  {author} {\bibfnamefont {A.}~\bibnamefont {Rubio}},\ }\href@noop {}
  {\bibfield  {journal} {\bibinfo  {journal} {Eur. Phys. J. D}\ }\textbf
  {\bibinfo {volume} {28}},\ \bibinfo {pages} {211} (\bibinfo {year}
  {2004})}\BibitemShut {NoStop}%
\bibitem [{\citenamefont {Isborn}, \citenamefont {Li},\ and\ \citenamefont
  {Tully}(2007)}]{Isborn_Tully:2007}%
  \BibitemOpen
  \bibfield  {author} {\bibinfo {author} {\bibfnamefont {C.~M.}\ \bibnamefont
  {Isborn}}, \bibinfo {author} {\bibfnamefont {X.}~\bibnamefont {Li}},\ and\
  \bibinfo {author} {\bibfnamefont {J.~C.}\ \bibnamefont {Tully}},\ }\href
  {https://doi.org/10.1063/1.2713391} {\bibfield  {journal} {\bibinfo
  {journal} {J.~Chem.\ Phys.}\ }\textbf {\bibinfo {volume} {126}},\ \bibinfo
  {pages} {134307} (\bibinfo {year} {2007})}\BibitemShut {NoStop}%
\bibitem [{\citenamefont {Miyamoto}, \citenamefont {Rubio},\ and\ \citenamefont
  {Tom\'anek}(2006)}]{Miyamoto_Tomanek:2006}%
  \BibitemOpen
  \bibfield  {author} {\bibinfo {author} {\bibfnamefont {Y.}~\bibnamefont
  {Miyamoto}}, \bibinfo {author} {\bibfnamefont {A.}~\bibnamefont {Rubio}},\
  and\ \bibinfo {author} {\bibfnamefont {D.}~\bibnamefont {Tom\'anek}},\ }\href
  {https://doi.org/10.1103/PhysRevLett.97.126104} {\bibfield  {journal}
  {\bibinfo  {journal} {Phys. Rev. Lett.}\ }\textbf {\bibinfo {volume} {97}},\
  \bibinfo {pages} {126104} (\bibinfo {year} {2006})}\BibitemShut {NoStop}%
\bibitem [{\citenamefont {Meng}\ and\ \citenamefont
  {Kaxiras}(2008)}]{Meng_Kaxiras:2008}%
  \BibitemOpen
  \bibfield  {author} {\bibinfo {author} {\bibfnamefont {S.}~\bibnamefont
  {Meng}}\ and\ \bibinfo {author} {\bibfnamefont {E.}~\bibnamefont {Kaxiras}},\
  }\href {https://doi.org/10.1063/1.2960628} {\bibfield  {journal} {\bibinfo
  {journal} {J.~Chem.\ Phys.}\ }\textbf {\bibinfo {volume} {129}},\ \bibinfo
  {pages} {054110} (\bibinfo {year} {2008})}\BibitemShut {NoStop}%
\bibitem [{\citenamefont {Andrade}\ \emph {et~al.}(2009)\citenamefont
  {Andrade}, \citenamefont {Castro}, \citenamefont {Zueco}, \citenamefont
  {Alonso}, \citenamefont {Echenique}, \citenamefont {Falceto},\ and\
  \citenamefont {Rubio}}]{Andrade_Rubio:2009}%
  \BibitemOpen
  \bibfield  {author} {\bibinfo {author} {\bibfnamefont {X.}~\bibnamefont
  {Andrade}}, \bibinfo {author} {\bibfnamefont {A.}~\bibnamefont {Castro}},
  \bibinfo {author} {\bibfnamefont {D.}~\bibnamefont {Zueco}}, \bibinfo
  {author} {\bibfnamefont {J.~L.}\ \bibnamefont {Alonso}}, \bibinfo {author}
  {\bibfnamefont {P.}~\bibnamefont {Echenique}}, \bibinfo {author}
  {\bibfnamefont {F.}~\bibnamefont {Falceto}},\ and\ \bibinfo {author}
  {\bibfnamefont {{\'A}.}~\bibnamefont {Rubio}},\ }\href
  {https://doi.org/10.1021/ct800518j} {\bibfield  {journal} {\bibinfo
  {journal} {J.~Chem.\ Theory Comput.}\ }\textbf {\bibinfo {volume} {5}},\
  \bibinfo {pages} {728} (\bibinfo {year} {2009})}\BibitemShut {NoStop}%
\bibitem [{\citenamefont {Liang}\ \emph {et~al.}(2010)\citenamefont {Liang},
  \citenamefont {Isborn}, \citenamefont {Lindsay}, \citenamefont {Li},
  \citenamefont {Smith},\ and\ \citenamefont {Levis}}]{Liang_Levis:2010}%
  \BibitemOpen
  \bibfield  {author} {\bibinfo {author} {\bibfnamefont {W.}~\bibnamefont
  {Liang}}, \bibinfo {author} {\bibfnamefont {C.~M.}\ \bibnamefont {Isborn}},
  \bibinfo {author} {\bibfnamefont {A.}~\bibnamefont {Lindsay}}, \bibinfo
  {author} {\bibfnamefont {X.}~\bibnamefont {Li}}, \bibinfo {author}
  {\bibfnamefont {S.~M.}\ \bibnamefont {Smith}},\ and\ \bibinfo {author}
  {\bibfnamefont {R.~J.}\ \bibnamefont {Levis}},\ }\href@noop {} {\bibfield
  {journal} {\bibinfo  {journal} {J.~Phys.\ Chem.~A}\ }\textbf {\bibinfo
  {volume} {114}},\ \bibinfo {pages} {6201} (\bibinfo {year}
  {2010})}\BibitemShut {NoStop}%
\bibitem [{\citenamefont {Castro}, \citenamefont {Marques},\ and\ \citenamefont
  {Rubio}(2004)}]{Castro_Rubio:2004a}%
  \BibitemOpen
  \bibfield  {author} {\bibinfo {author} {\bibfnamefont {A.}~\bibnamefont
  {Castro}}, \bibinfo {author} {\bibfnamefont {M.~A.~L.}\ \bibnamefont
  {Marques}},\ and\ \bibinfo {author} {\bibfnamefont {A.}~\bibnamefont
  {Rubio}},\ }\href {https://doi.org/10.1063/1.1774980} {\bibfield  {journal}
  {\bibinfo  {journal} {J.~Chem.\ Phys.}\ }\textbf {\bibinfo {volume} {121}},\
  \bibinfo {pages} {3425} (\bibinfo {year} {2004})}\BibitemShut {NoStop}%
\bibitem [{\citenamefont {Marques}\ and\ \citenamefont
  {Gross}(2003)}]{book_Marques_Gross:2003}%
  \BibitemOpen
  \bibfield  {author} {\bibinfo {author} {\bibfnamefont {M.~A.~L.}\
  \bibnamefont {Marques}}\ and\ \bibinfo {author} {\bibfnamefont {E.~K.~U.}\
  \bibnamefont {Gross}},\ }\enquote {\bibinfo {title} {Time-dependent density
  functional theory},}\ in\ \href {https://doi.org/10.1007/3-540-37072-2_4}
  {\emph {\bibinfo {booktitle} {A Primer in Density Functional Theory}}},\
  \bibinfo {editor} {edited by\ \bibinfo {editor} {\bibfnamefont
  {C.}~\bibnamefont {Fiolhais}}, \bibinfo {editor} {\bibfnamefont
  {F.}~\bibnamefont {Nogueira}},\ and\ \bibinfo {editor} {\bibfnamefont
  {M.~A.~L.}\ \bibnamefont {Marques}}}\ (\bibinfo  {publisher} {Springer Berlin
  Heidelberg},\ \bibinfo {address} {Berlin, Heidelberg},\ \bibinfo {year}
  {2003})\ pp.\ \bibinfo {pages} {144--184}\BibitemShut {NoStop}%
\bibitem [{\citenamefont {Marques}\ \emph {et~al.}(2003)\citenamefont
  {Marques}, \citenamefont {Castro}, \citenamefont {Bertsch},\ and\
  \citenamefont {Rubio}}]{Marques_Rubio:2003}%
  \BibitemOpen
  \bibfield  {author} {\bibinfo {author} {\bibfnamefont {M.~A.}\ \bibnamefont
  {Marques}}, \bibinfo {author} {\bibfnamefont {A.}~\bibnamefont {Castro}},
  \bibinfo {author} {\bibfnamefont {G.~F.}\ \bibnamefont {Bertsch}},\ and\
  \bibinfo {author} {\bibfnamefont {A.}~\bibnamefont {Rubio}},\ }\href
  {https://doi.org/https://doi.org/10.1016/S0010-4655(02)00686-0} {\bibfield
  {journal} {\bibinfo  {journal} {Comput. Phys. Commun}\ }\textbf {\bibinfo
  {volume} {151}},\ \bibinfo {pages} {60} (\bibinfo {year} {2003})}\BibitemShut
  {NoStop}%
\bibitem [{\citenamefont {Andrade}\ \emph {et~al.}(2012)\citenamefont
  {Andrade}, \citenamefont {Alberdi-Rodriguez}, \citenamefont {Strubbe},
  \citenamefont {Oliveira}, \citenamefont {Nogueira}, \citenamefont {Castro},
  \citenamefont {Muguerza}, \citenamefont {Arruabarrena}, \citenamefont
  {Louie}, \citenamefont {Aspuru-Guzik}, \citenamefont {Rubio},\ and\
  \citenamefont {Marques}}]{Andrade_Marques:2012}%
  \BibitemOpen
  \bibfield  {author} {\bibinfo {author} {\bibfnamefont {X.}~\bibnamefont
  {Andrade}}, \bibinfo {author} {\bibfnamefont {J.}~\bibnamefont
  {Alberdi-Rodriguez}}, \bibinfo {author} {\bibfnamefont {D.~A.}\ \bibnamefont
  {Strubbe}}, \bibinfo {author} {\bibfnamefont {M.~J.~T.}\ \bibnamefont
  {Oliveira}}, \bibinfo {author} {\bibfnamefont {F.}~\bibnamefont {Nogueira}},
  \bibinfo {author} {\bibfnamefont {A.}~\bibnamefont {Castro}}, \bibinfo
  {author} {\bibfnamefont {J.}~\bibnamefont {Muguerza}}, \bibinfo {author}
  {\bibfnamefont {A.}~\bibnamefont {Arruabarrena}}, \bibinfo {author}
  {\bibfnamefont {S.~G.}\ \bibnamefont {Louie}}, \bibinfo {author}
  {\bibfnamefont {A.}~\bibnamefont {Aspuru-Guzik}}, \bibinfo {author}
  {\bibfnamefont {A.}~\bibnamefont {Rubio}},\ and\ \bibinfo {author}
  {\bibfnamefont {M.~A.~L.}\ \bibnamefont {Marques}},\ }\href
  {https://doi.org/10.1088/0953-8984/24/23/233202} {\bibfield  {journal}
  {\bibinfo  {journal} {J.~Phys.: Cond.\ Matt.}\ }\textbf {\bibinfo {volume}
  {24}},\ \bibinfo {pages} {233202} (\bibinfo {year} {2012})}\BibitemShut
  {NoStop}%
\bibitem [{\citenamefont {Tancogne-Dejean}\ \emph {et~al.}(2020)\citenamefont
  {Tancogne-Dejean}, \citenamefont {Oliveira}, \citenamefont {Andrade},
  \citenamefont {Appel}, \citenamefont {Borca}, \citenamefont {Le~Breton},
  \citenamefont {Buchholz}, \citenamefont {Castro}, \citenamefont {Corni},
  \citenamefont {Correa}, \citenamefont {De~Giovannini}, \citenamefont
  {Delgado}, \citenamefont {Eich}, \citenamefont {Flick}, \citenamefont {Gil},
  \citenamefont {Gomez}, \citenamefont {Helbig}, \citenamefont {H{\"u}bener},
  \citenamefont {Jest{\"a}dt}, \citenamefont {Jornet-Somoza}, \citenamefont
  {Larsen}, \citenamefont {Lebedeva}, \citenamefont {L{\"u}ders}, \citenamefont
  {Marques}, \citenamefont {Ohlmann}, \citenamefont {Pipolo}, \citenamefont
  {Rampp}, \citenamefont {Rozzi}, \citenamefont {Strubbe}, \citenamefont
  {Sato}, \citenamefont {Sch{\"a}fer}, \citenamefont {Theophilou},
  \citenamefont {Welden},\ and\ \citenamefont
  {Rubio}}]{Tancogne-Dejean_Rubio:2020}%
  \BibitemOpen
  \bibfield  {author} {\bibinfo {author} {\bibfnamefont {N.}~\bibnamefont
  {Tancogne-Dejean}}, \bibinfo {author} {\bibfnamefont {M.~J.~T.}\ \bibnamefont
  {Oliveira}}, \bibinfo {author} {\bibfnamefont {X.}~\bibnamefont {Andrade}},
  \bibinfo {author} {\bibfnamefont {H.}~\bibnamefont {Appel}}, \bibinfo
  {author} {\bibfnamefont {C.~H.}\ \bibnamefont {Borca}}, \bibinfo {author}
  {\bibfnamefont {G.}~\bibnamefont {Le~Breton}}, \bibinfo {author}
  {\bibfnamefont {F.}~\bibnamefont {Buchholz}}, \bibinfo {author}
  {\bibfnamefont {A.}~\bibnamefont {Castro}}, \bibinfo {author} {\bibfnamefont
  {S.}~\bibnamefont {Corni}}, \bibinfo {author} {\bibfnamefont {A.~A.}\
  \bibnamefont {Correa}}, \bibinfo {author} {\bibfnamefont {U.}~\bibnamefont
  {De~Giovannini}}, \bibinfo {author} {\bibfnamefont {A.}~\bibnamefont
  {Delgado}}, \bibinfo {author} {\bibfnamefont {F.~G.}\ \bibnamefont {Eich}},
  \bibinfo {author} {\bibfnamefont {J.}~\bibnamefont {Flick}}, \bibinfo
  {author} {\bibfnamefont {G.}~\bibnamefont {Gil}}, \bibinfo {author}
  {\bibfnamefont {A.}~\bibnamefont {Gomez}}, \bibinfo {author} {\bibfnamefont
  {N.}~\bibnamefont {Helbig}}, \bibinfo {author} {\bibfnamefont
  {H.}~\bibnamefont {H{\"u}bener}}, \bibinfo {author} {\bibfnamefont
  {R.}~\bibnamefont {Jest{\"a}dt}}, \bibinfo {author} {\bibfnamefont
  {J.}~\bibnamefont {Jornet-Somoza}}, \bibinfo {author} {\bibfnamefont {A.~H.}\
  \bibnamefont {Larsen}}, \bibinfo {author} {\bibfnamefont {I.~V.}\
  \bibnamefont {Lebedeva}}, \bibinfo {author} {\bibfnamefont {M.}~\bibnamefont
  {L{\"u}ders}}, \bibinfo {author} {\bibfnamefont {M.~A.~L.}\ \bibnamefont
  {Marques}}, \bibinfo {author} {\bibfnamefont {S.~T.}\ \bibnamefont
  {Ohlmann}}, \bibinfo {author} {\bibfnamefont {S.}~\bibnamefont {Pipolo}},
  \bibinfo {author} {\bibfnamefont {M.}~\bibnamefont {Rampp}}, \bibinfo
  {author} {\bibfnamefont {C.~A.}\ \bibnamefont {Rozzi}}, \bibinfo {author}
  {\bibfnamefont {D.~A.}\ \bibnamefont {Strubbe}}, \bibinfo {author}
  {\bibfnamefont {S.~A.}\ \bibnamefont {Sato}}, \bibinfo {author}
  {\bibfnamefont {C.}~\bibnamefont {Sch{\"a}fer}}, \bibinfo {author}
  {\bibfnamefont {I.}~\bibnamefont {Theophilou}}, \bibinfo {author}
  {\bibfnamefont {A.}~\bibnamefont {Welden}},\ and\ \bibinfo {author}
  {\bibfnamefont {A.}~\bibnamefont {Rubio}},\ }\href
  {https://doi.org/10.1063/1.5142502} {\bibfield  {journal} {\bibinfo
  {journal} {J.~Chem.\ Phys.}\ }\textbf {\bibinfo {volume} {152}},\ \bibinfo
  {pages} {124119} (\bibinfo {year} {2020})}\BibitemShut {NoStop}%
\bibitem [{\citenamefont {Lee}(2009)}]{book_Lee:2009}%
  \BibitemOpen
  \bibfield  {author} {\bibinfo {author} {\bibfnamefont {J.~M.}\ \bibnamefont
  {Lee}},\ }\href@noop {} {\emph {\bibinfo {title} {Manifolds and Differential
  Geometry}}},\ Vol.\ \bibinfo {volume} {107}\ (\bibinfo  {publisher} {American
  Mathematical Soc.},\ \bibinfo {year} {2009})\BibitemShut {NoStop}%
\bibitem [{\citenamefont {Kosloff}\ and\ \citenamefont
  {Tal-Ezer}(1986)}]{Kosloff_Tal-Ezer:1986}%
  \BibitemOpen
  \bibfield  {author} {\bibinfo {author} {\bibfnamefont {R.}~\bibnamefont
  {Kosloff}}\ and\ \bibinfo {author} {\bibfnamefont {H.}~\bibnamefont
  {Tal-Ezer}},\ }\href {https://doi.org/10.1016/0009-2614(86)80262-7}
  {\bibfield  {journal} {\bibinfo  {journal} {Chem.\ Phys.\ Lett.}\ }\textbf
  {\bibinfo {volume} {127}},\ \bibinfo {pages} {223} (\bibinfo {year}
  {1986})}\BibitemShut {NoStop}%
\end{thebibliography}%

\end{document}